\documentclass[12pt]{amsart}
\usepackage{amsmath,amssymb,cite,enumerate,graphicx}
\usepackage[foot]{amsaddr}
\usepackage[sc]{mathpazo}

\numberwithin{equation}{section}

\newcommand\BS{\boldsymbol}
\newcommand\dif{\,\mathrm{d}}
\newcommand\tr{\mathrm{Tr}}
\newcommand\deriv[2]{\frac{\dif #1}{\dif #2}}
\newcommand\parderiv[2]{\frac{\partial #1}{\partial #2}}
\newcommand\coll{\mathcal C}
\renewcommand\div{\mathrm{div}}
\newcommand\Pran{\mbox{\textit{Pr}}}

\newcommand\symtr[1]{\big[ #1 \big]_{TS}}

\newtheorem{proposition}{Proposition}

\begin{document}

\author[Rafail V. Abramov]{Rafail V. Abramov}

\address{Department of Mathematics, Statistics and Computer Science,
University of Illinois at Chicago, 851 S. Morgan st., Chicago, IL 60607}

\email{abramov@math.uic.edu}

\title[Coarse-grained transport via moments of averaged Boltzmann
  equation]{Coarse-grained transport of a turbulent flow via moments
  of the Reynolds-averaged Boltzmann equation}

\begin{abstract}
Here we introduce new coarse-grained variables for a turbulent flow in
the form of moments of its Reynolds-averaged Boltzmann equation. With
the exception of the collision moments, the transport equations for
the new variables are identical to the usual moment equations, and
thus naturally lend themselves to the variety of already existing
closure methods. Under the anelastic turbulence approximation, we
derive equations for the Reynolds-averaged turbulent fluctuations
around the coarse-grained state. We show that the global relative
entropy of the coarse-grained state is bounded from above by the
Reynolds average of the fine-grained global relative entropy, and thus
obeys the time decay bound of Desvillettes and Villani. This is
similar to what is observed in the rarefied gas dynamics, which makes
the Grad moment closure a good candidate for truncating the hierarchy
of the coarse-grained moment equations. We also show that, under
additional assumptions on the form of the coarse-grained collision
terms, one arrives at the Navier-Stokes closure, which can be
naturally extended to the Burnett and super-Burnett orders. Finally,
we suggest crude parameterizations of the coarse-grained collision
terms for use as starting points in numerical simulation and modeling.
\end{abstract}

\date{\today}

\maketitle

\section{Introduction}

A common principle in turbulence modeling involves the averaging of
the well-known Navier-Stokes equations to filter out the rapid
small-scale turbulent fluctuations from their solutions to reduce
computational cost. This is usually done via the Reynolds averaging of
the Navier-Stokes equations \cite{Rey,MonYag,MonYag2,Wilc}, which
produces additional nonlinear terms in the equations. These additional
terms are often modeled via the turbulent eddy viscosity assumption,
which was suggested back in 1877 by Boussinesq \cite{Bou}, and which
generally provides good agreement with experiments in near-wall
boundary layers.

However, for the fully developed three-dimensional turbulence it was
observed that the concept of the turbulent eddy viscosity fails
\cite{Kay}. Girimaji \cite{Gir} argued that the reason why the
turbulent viscosity approximation could be invalid for the averaged
Navier-Stokes equations was that the latter required that the
corresponding solution of the Boltzmann equation
\cite{Cer,Cer2,Lev,Gols} was near its Maxwellian equilibrium, while
the corresponding averaged solution did not have to be near
equilibrium. As an alternative to the Reynolds averaging of the
Navier-Stokes equations, Girimaji \cite{Gir} proposed the direct
filtering of the solution of the Boltzmann equation instead, and then
solve the filtered equation using the lattice Boltzmann method (LBM).

However, Girimaji's approach had a few drawbacks. First, the Reynolds
stress closure problem was not eliminated in \cite{Gir}, and the
Reynolds stress was modeled by the standard Smagorinsky-Lilly closure
\cite{Smag}, the assumptions of which can be traced back to the
Boussinesq approximation \cite{Bou}. Second, the Bhatnagar-Gross-Krook
(BGK) approximation of the collision terms \cite{BhaGroKro}, which was
used in \cite{Gir}, sets the Prandtl number of the flow strictly to 1,
whereas it is known that the Prandtl number of a fluid is generally
different from 1; for example, for a monatomic ideal gas it equals
$2/3$, and its value for the air is around 0.7-0.8 (there are,
however, improved BGK collision parameterizations with non-unitary
Prandtl numbers \cite{AndPer,AndTalPerPer}). Third, the lattice
Boltzmann method usually involves more computational variables than
the standard fluid dynamics methods, which may limit its use in some
applications.

In this work we propose a new coarse-graining approach were the
Reynolds averaging is used directly on the Boltzmann equation like in
\cite{Gir}, however, we further convert the resulting
Reynolds-averaged Boltzmann equation into the hierarchy of the
coarse-grained moment equations. While the new coarse-grained moments
are different from the usual Reynolds averages of moment variables
used in conventional transport methods, the resulting hierarchy of the
transport equations for the coarse-grained moments is identical to the
usual moment transport hierarchy, with the exception of the nonlinear
collision terms. We show that the global relative entropy of the
Reynolds-averaged Boltzmann equation is bounded from above by the
Reynolds average of the global relative entropy of the usual Boltzmann
equation, and, therefore, obeys the same bound on the decay rate as
established by Desvillettes and Villani \cite{DesVil} for the usual
Boltzmann equation. This justifies the same closure methodologies as
are used to truncate the moment transport equations for the rarefied
gas dynamics \cite{Cer3}, in particular the Grad \cite{Gra,Gra2}
closure.

For the modeling of statistical properties of the turbulent
fluctuations, we derive the equation for the transport of the
turbulent kinetic energy under the approximation of the anelastic
turbulence, as well as the appropriate relations for the turbulent
components of the stress and heat flux. There remain three unspecified
dissipation terms: the turbulent energy dissipation rate, and the two
collision moments for the coarse-grained stress and heat flux. The
reason why the coarse-grained collision terms remain unspecified is
because they originate from the Reynolds average of the nonlinear
collision operator of the Boltzmann equation, and thus require
additional information about the structure of the statistical ensemble
(and, therefore, the physics of the flow). We suggest crude
approximations for these remaining nonlinear terms, based on a
dimensional reasoning.

The manuscript is organized as follows. In Section
\ref{sec:conventional_moment_closures} we discuss the Boltzmann
equation and its conventional closures: the Euler, Navier-Stokes and
Grad equations. In Section \ref{sec:coarse_grained} we introduce the
Reynolds averaging operator, define the new coarse-grained variables,
and show that the transport equations for the new coarse-grained
variables have the same hierarchy as those for the conventional
fine-grained variables, with different collision terms. In Section
\ref{sec:turbulent_fluctuations} we parameterize the turbulent
fluctuations of the stress and heat flux, and derive the transport
equation for the turbulent energy. In Section
\ref{sec:coarse_grained_moment_closures} we describe the Grad and
Navier-Stokes moment closures for the new coarse-grained transport
equations, as well as mention how to extend the latter to the Burnett
and super-Burnett orders. In Section
\ref{sec:computational_considerations} we discuss basic approaches to
the modeling of the coarse-grained collision terms and the turbulent
energy dissipation rate. In Section \ref{sec:summary} we summarize the
results and discuss future work.

\section{Conventional moment closures of the Boltzmann equation}
\label{sec:conventional_moment_closures}

In the absence of external forces, the Boltzmann equation for a
3-dimensional fluid is given by \cite{Cer,Cer2}
\begin{equation}
\label{eq:boltzmann_equation}
\parderiv ft+\BS v\cdot\nabla_{\BS x}f=\coll(f).
\end{equation}
Here, $t$ and $\BS x$ are the time and space coordinates, $\BS v$ is
the velocity of a fluid particle, and $f(t,\BS x,\BS v)$ is the
statistical velocity distribution of the fluid particles, at the
location $\BS x$ and time $t$. The left-hand side contains the
transport terms for $f$, while the right-hand side contains the
nonlinear {\em collision term}, which generally has the effect of
dissipation. In many applications, $\coll(f)$ is assumed to be
bilinear in $f$, as the situations where three or more particles
collide at once rarely occur.

Let the angle brackets denote the average over the fluid particles
$\BS v$:
\begin{subequations}
\begin{equation}
\langle b\rangle_f(t,\BS x)=\int b(\BS v)f(t,\BS x,\BS v)\dif\BS v,
\end{equation}
\begin{equation}
\langle b\rangle_{\coll(f)}(t,\BS x)=\int b(\BS v)\coll(f(t,\BS x,\BS
v))\dif\BS v.
\end{equation}
\end{subequations}
for an integrable function $b(\BS v)$. Then, one can apply these
averages onto the Boltzmann equation in \eqref{eq:boltzmann_equation}
and obtain the corresponding moment transport equation:
\begin{equation}
\label{eq:moment_transport_equation}
\parderiv{}t\langle b\rangle_f+\div_{\BS x}\langle b\BS
v\rangle_f=\langle b \rangle_{\coll(f)}.
\end{equation}
The nonlinear collision term $\coll(f)$ has the requirement of the
mass, momentum, and energy conservation:
\begin{equation}
\label{eq:collision_conservation_laws}
\langle 1\rangle_{\coll(f)}=0, \qquad
\langle\BS v\rangle_{\coll(f)}=\BS 0, \qquad
\langle\|\BS v\|^2\rangle_{\coll(f)}=0.
\end{equation}
The last identity also signifies that there is no ``internal energy''
in the fluid -- all energy that the fluid carries is confined to the
velocity of its particles. This transport-collision model applies, for
example, to a monatomic ideal gas. Here we adopt this model for the
sake of simplicity of illustration, as the presence of ``hidden''
energy-accumulating degrees of freedom in the fluid particles requires
a different, more complicated treatment of the Boltzmann equation,
which will be presented elsewhere.

In order to obtain the conventional moment closures from the moment
transport equation \eqref{eq:moment_transport_equation}, we first
introduce the following conventional velocity moments of $f$:
\begin{subequations}
\label{eq:moments}
\begin{equation}
\label{eq:density}
\rho=\langle 1\rangle_f,\qquad\text{density},
\end{equation}
\begin{equation}
\label{eq:momentum}
\rho\BS u=\langle\BS v\rangle_f,\qquad\text{momentum},
\end{equation}
\begin{equation}
\label{eq:pressure}
p=\frac 13\langle\|\BS v-\BS u\|^2\rangle_f,\qquad\text{pressure},
\end{equation}
\begin{equation}
\label{eq:stress}
\BS S=\langle(\BS v-\BS u)\otimes(\BS v-\BS u)\rangle_f-p\BS I,
\qquad\text{stress},
\end{equation}
\begin{equation}
\label{eq:heat_flux}
\BS q=\frac 12\langle\|\BS v-\BS u\|^2(\BS v-\BS u)\rangle_f,
\qquad\text{heat flux},
\end{equation}
\end{subequations}
where ``$\otimes$'' denotes the outer product of two vectors.
Rearranging the averages \eqref{eq:pressure}--\eqref{eq:heat_flux}
above, one writes the identities
\begin{subequations}
\begin{equation}
\label{eq:pressure_2}
\langle\|\BS v\|^2\rangle_f=3p+\rho\|\BS u\|^2,
\end{equation}
\begin{equation}
\langle\BS v\otimes\BS v\rangle_f=\BS S+p\BS I+\rho\BS u\otimes\BS
u,
\end{equation}
\begin{equation}
\label{eq:heat_flux_2}
\frac 12\langle\|\BS v\|^2\BS v\rangle_f=\BS q+\BS S\BS u+\frac 52
p\BS u+\frac 12\rho\|\BS u\|^2\BS u.
\end{equation}
\end{subequations}
Now, writing the moment transport equation in
\eqref{eq:moment_transport_equation} for the velocity moments in
\eqref{eq:density}, \eqref{eq:momentum},
\eqref{eq:pressure_2}--\eqref{eq:heat_flux_2}, expressing those
moments in terms of $\rho$, $\BS u$, $p$, $\BS S$ and $\BS q$
using the identities above, and taking into account the mass,
momentum, and energy conservation laws in
\eqref{eq:collision_conservation_laws}, one arrives at the following
system of equations:
\begin{subequations}
\label{eq:transport_equations}
\begin{equation}
\label{eq:density_equation}
\parderiv\rho t+\div(\rho\BS u)=0,
\end{equation}
\begin{equation}
\label{eq:momentum_equation}
\parderiv{(\rho\BS u)}t+\div(\rho\BS u\otimes\BS u+p\BS I+\BS S)
=\BS 0,
\end{equation}
\begin{equation}
\label{eq:pressure_equation}
\parderiv pt+\div(p\BS u)+\frac 23(p\,\div\BS u+\BS S:(\nabla
\otimes\BS u)+\div\BS q)=0,
\end{equation}
\begin{equation}
\label{eq:stress_equation}
\parderiv{\BS S}t+(\BS u\cdot\nabla)\BS S+\div(\BS u)\BS S
+2\symtr{\BS S(\nabla\otimes\BS u)}+\div\BS Q-\frac 23(\div \BS
q)\BS I+2p\symtr{\nabla\otimes\BS u}=\BS C_S,
\end{equation}
\begin{equation}
\label{eq:heat_flux_equation}
\parderiv{\BS q}t+\div(\BS q\otimes\BS u)+(\BS q\cdot\nabla)\BS u-
\frac 1\rho\left(\frac 52p\BS I+\BS S\right)\div(p\BS I+\BS S)
+\BS Q:(\nabla\otimes\BS u)+\div\BS R=\BS c_q,
\end{equation}
\end{subequations}
where
\begin{equation}
\BS C_S=\langle \BS v\otimes\BS v\rangle_{\coll(f)},\qquad
\BS c_q=\frac 12\langle(\|\BS v\|^2-2\BS u\cdot\BS v)\BS
v\rangle_{\coll(f)},
\end{equation}
the symbol ``$:$'' denotes the Frobenius product of two matrices, and
$\symtr{\BS A}$ denotes the traceless symmetrization of a $3\times 3$
matrix $\BS A$:
\begin{equation}
\symtr{\BS A}=\frac 12\left(\BS A+\BS A^T\right)-\frac 13{\tr(\BS
  A)}\BS I.
\end{equation}
Above in \eqref{eq:transport_equations}, $\BS Q$ and $\BS R$ are the
unknown higher-order moments,
\begin{subequations}
\begin{equation}
\BS Q=\langle(\BS v-\BS u)\otimes(\BS v-\BS u)\otimes(\BS v-\BS u)
\rangle_f
\end{equation}
being the full 3-rank skewness tensor, and
\begin{equation}
\BS R=\frac 12\langle\|\BS v-\BS u\|^2(\BS v-\BS u)\otimes(\BS v-\BS
u)\rangle_f
\end{equation}
\end{subequations}
being the matrix of the contracted 4th-order moment. Both $\BS Q$ and
$\BS R$ obey their own transport equations, which, in turn, obviously
include moments of yet higher orders, and so forth.

\begin{figure}
\includegraphics[width=0.6\textwidth]{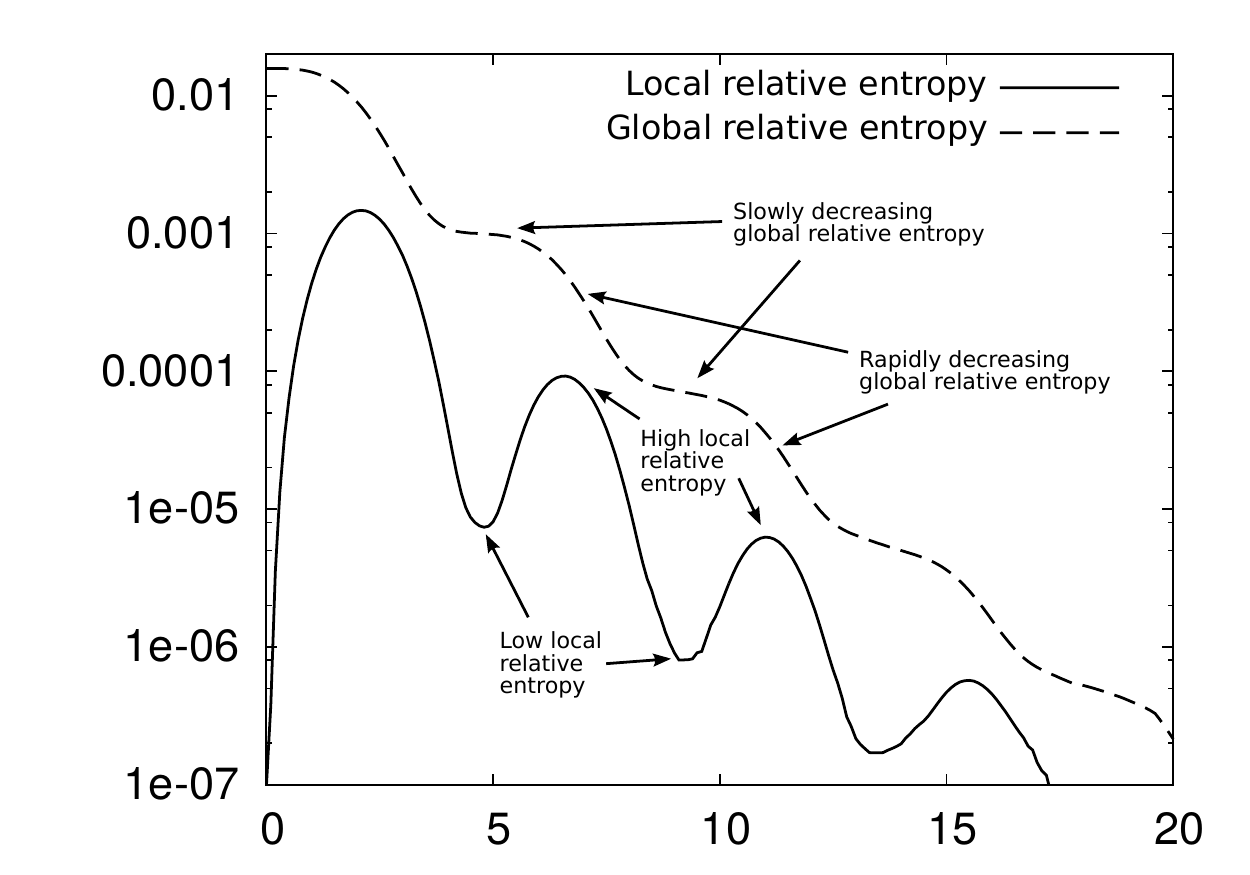}
\caption{Time series of the local and global relative entropies of a
  solution of the Boltzmann equation. This plot is adapted from Figure
  8, upper-right plot, of the article ``Solving the Boltzmann equation
  in ${N}\log_2{N}$'' by F.~Filbet, C.~Mouhot, and L.~Pareschi,
  published in {\em SIAM J. Sci. Comp.}, 28(3):1029--1053, 2006 (see
  reference \cite{FilMouPar} in the bibliography section). Used with
  permission of F.~Filbet, C.~Mouhot, L.~Pareschi and the Society for
  Industrial and Applied Mathematics.}
\label{fig:entropies}
\end{figure}

In order to understand how to close the moment equations above, we
turn to the well-known Boltzmann's {\em H}-theorem for ideal gases.
We state Boltzmann's {\em H}-theorem in the same form as in Golse
\cite{Gols}:
\begin{proposition}[Boltzmann's {\em H}-theorem]
The following inequality holds for the collision term:
\begin{equation}
\label{eq:boltzmann_inequality}
\langle\ln f\rangle_{\mathcal C(f)}\leq 0.
\end{equation}
Moreover, the following three conditions are equivalent:
\begin{enumerate}
\item $\langle\ln f\rangle_{\mathcal C(f)}=0$,
\item $\coll(f)=0$ for all $\BS v\in\mathbb R^3$,
\item $f$ is the local Maxwellian distribution,
\begin{equation}
\label{eq:Maxwellian}
f_M(\BS v)=\frac\rho{(2\pi\theta)^{3/2}}\exp\left(-\frac{\|\BS v-\BS
  u\|^2}{2\theta}\right),
\end{equation}
where $\theta$ denotes the temperature
\begin{equation}
\label{eq:temperature}
\theta=\frac p\rho.
\end{equation}
\end{enumerate}
\end{proposition}
In order to make use of this theorem, one introduces the local and
global entropy functionals as follows. The local entropy
$S_l[f](t,\BS x)$ is given by the functional
\begin{equation}
\label{eq:entropy}
S_l[f](t,\BS x)=-\int f\ln f\dif\BS v=-\langle\ln f\rangle_f.
\end{equation}
The global entropy $S_g[f](t)$ is further given by
\begin{equation}
\label{eq:global_entropy}
S_g[f](t)=\int S_l[f]\dif\BS x=-\int\langle\ln f\rangle_f\dif x.
\end{equation}
Now, we can look at the evolution equation for $\langle\ln
f\rangle_f$, which is given by
\begin{equation}
\parderiv{}t\langle\ln f\rangle_f=\int (\ln f+1)\parderiv ft\dif\BS
v=\int (\ln f+1)\left(-\BS v\cdot\nabla_{\BS x}f+\mathcal
C(f)\right)\dif\BS v.
\end{equation}
Observing the mass conservation law in
\eqref{eq:collision_conservation_laws}, we further derive
\begin{equation}
\label{eq:local_entropy_equation}
\parderiv{}t\langle\ln f\rangle_f+\div_{\BS x}\langle\BS v\ln
f\rangle_f=\langle\ln f\rangle_{\mathcal C(f)}.
\end{equation}
Assuming that one can omit the effect of the spatial boundaries when
integrating, we obtain for the global entropy
\begin{equation}
\label{eq:global_entropy_equation}
\deriv{}tS_g[f]=-\int\langle\ln f\rangle_{\mathcal C(f)}\dif\BS x.
\end{equation}
Qualitatively, the equations in \eqref{eq:local_entropy_equation} and
\eqref{eq:global_entropy_equation} do, roughly, the following:
\begin{enumerate}
\item The equation for the local entropy in
  \eqref{eq:local_entropy_equation} tends to increase the local
  entropy $S_l[f]$, unless $f$ is already the local Maxwellian
  state in \eqref{eq:Maxwellian}, in which case $S_l[f]$ is already at
  its maximum for given local constraints $\rho$, $\BS u$ and $\theta$
  (or, equivalently, $p$). The rate of convergence towards the local
  maximum entropy state here is usually very rapid, however, the
  presence of advection prevents $S_l[f]$ from reaching its local
  maximum at \eqref{eq:Maxwellian}.
\item The equation for the global entropy in
  \eqref{eq:global_entropy_equation} tends to increase $S_g[f]$,
  unless $f$ is the global Maxwellian distribution of the form
\begin{equation}
\label{eq:global_Maxwellian}
f_M^{\text{g}}(\BS v)=\frac{\rho_0}{(2\pi\theta_0)^{3/2}}
\exp\left(-\frac{\|\BS v-\BS u_0\|^2}{2\theta_0}\right),
\end{equation}
where $\rho_0$, $\BS u_0$ and $\theta_0$ are constants throughout the
spatial domain, specified by the total mass, momentum and energy
constraints in the system. This process is unaffected by advection,
and the time rate of convergence to the global maximum state is
$O(t^{-\infty})$ (Desvillettes and Villani \cite{DesVil}), that is,
rather slow.
\end{enumerate}
These two processes are, in a certain sense, ``mutually exclusive'',
that is, the closer the $S[f](t,\BS x)$ is to its local maximums at
each $\BS x$, the slower the rate with which $S_g[f](t)$ approaches
its global maximum. This relation between the two processes can be
quantified more systematically by introducing the relative entropy
(Kullback-Leibler distance, \cite{KulLei}) $H[f|g]$ between two
distributions $f$ and $g$ as
\begin{equation}
H[f|g](t)=\int f\ln(f/g)\dif\BS v\dif\BS x.
\end{equation}
It is easy to show that $H[f|g]$ is always non-negative, and is zero
if and only if $f=g$. Now we denote the local and global relative
entropies as
\begin{equation}
H_l[f]=H\left[f\left|f_M\right.\right],\qquad
H_g[f]=H\left[f\left|f_M^{\text{g}}\right.\right],
\end{equation}
respectively. Then, $H_l[f]$ and $H_g[f]$ measure the
information-theoretic distance between $f$ and \eqref{eq:Maxwellian}
or \eqref{eq:global_Maxwellian}, respectively, becoming quantitative
indicators of the relation between the local and global entropy
processes, described above. The ``mutually exclusive'' behavior (slow
decrease rate of $H_g$ for small values of $H_l$, and, vice-versa,
rapid decrease of $H_g$ for large values of $H_l$) can be observed in
Figure 5 of \cite{DesVil}, and also in Figures 8 and 9 of Filbet,
Mouhot and Pareschi \cite{FilMouPar}, where the time series of the
local and global Kullback-Leibler distance between $f$ and both
\eqref{eq:Maxwellian} and \eqref{eq:global_Maxwellian} are displayed
for a direct numerical simulation with a Boltzmann equation. In Figure
\ref{fig:entropies} we show an example of such time series, adapted
from Figure 8 of \cite{FilMouPar}.
In fact, if, in a hypothetical situation, $S_l[f]$ is always at its
maximum at all points $\BS x$ (which amounts to strictly zero
$H_l[f]$), then $S_g[f]$ must be constant in time even if it is below
its global possible maximum (or, equivalently, $H_g[f]$ must be
constant in time even if nonzero). This assumption is, of course,
unrealistic in practice, however, it is used in the Euler closure of
the moment equations as we show below.

Different assumptions about the time-scale separation between the
local and global entropy processes lead to three different
conventional moment closures of the moment transport equations in
\eqref{eq:transport_equations}.

\subsection{The Euler closure}

In the Euler closure, it is assumed that time-scale separation between
the local and global entropy processes is infinite. Namely, it is
assumed that the local entropy $S_l[f]$ always instantaneously jumps
to its maximum for given $\rho$, $\BS u$ and $\theta$, and $f(t,\BS
x,\BS v)$ is permanently the Maxwellian state in
\eqref{eq:Maxwellian}. The global entropy $S_g[f]$ is, therefore,
fixed at its initial value, irrespective of its maximum state for the
present total mass, momentum and energy constraints. This assumption
allows to set the stress $\BS S$ and heat flux $\BS q$ identically
to zeros everywhere, since their corresponding velocity moments are
indeed zeros for the Maxwellian distribution in
\eqref{eq:Maxwellian}. The resulting famous Euler equations are given
by
\begin{subequations}
\label{eq:Euler_equations}
\begin{equation}
\parderiv\rho t+\div(\rho\BS u)=0,
\end{equation}
\begin{equation}
\parderiv{(\rho\BS u)}t+\div(\rho\BS u\otimes\BS u)+\nabla p=\BS 0,
\end{equation}
\begin{equation}
\parderiv pt+\div(p\BS u)+\frac 23p\,\div\BS u=0.
\end{equation}
\end{subequations}
The practical application of the Euler equations is somewhat limited
due to the lack of any kind of diffusion.

\subsection{The Navier-Stokes closure}

The Navier-Stokes closure is more sophisticated than the Euler closure
in the sense that it recognizes the finiteness of the time scale of
the local entropy dissipation process, as well as takes into account
the correction in the local entropy process due to advection. As a
result, dissipation of the global entropy is present on the long time
scale, which is more physically realistic than the constant global
entropy in the Euler closure.

As a result of statistical analysis of collisions of large numbers of
ideal spheres (which ideal gas is modeled upon), one arrives at the
following first-order linear approximations of the stress and heat
flux collision operators in the equations for the stress
\eqref{eq:stress_equation} and heat flux
\eqref{eq:heat_flux_equation}:
\begin{equation}
\label{eq:Navier_Stokes_collisions}
\BS C_S\approx-\frac p\mu\BS S,\qquad\BS c_q\approx-\frac
52\frac p\kappa\BS q,
\end{equation}
where the coefficients $\mu$ and $\kappa$ are the {\em viscosity} and
{\em heat conductivity} of the fluid, respectively (for more details,
see, for example, \cite{Gols,Gra}). Both $\mu$ and $\kappa$ are
defined largely by the physical properties of colliding spheres, and
for ideal gases they also weakly depend on the temperature
$\theta$. The ratio of the heat flux over stress dissipation rates is
known as the Prandtl number,
\begin{equation}
\Pran=\frac{\text{heat flux dissipation rate}}{\text{stress
    dissipation rate}}=\frac 52\frac\mu\kappa,
\end{equation}
which for monatomic gases equals precisely $2/3$ \cite{Gra}.

In what follows, we prefer to keep both the viscosity $\mu$ and heat
conductivity $\kappa$ as unrelated parameters, in order to distinguish
between the terms related to the stress, and those related to the heat
flux. However, it must be noted that, strictly put, all derivations
below are only valid when $15\mu=4\kappa$, since this is what the
monatomic gas model inherently implies.

Both $\mu$ and $\kappa$ are proportional to the masses of colliding
spheres, which, in the case of molecules, renders both $\mu$ and
$\kappa$ very small. As a result, the time scale of local entropy
dissipation is $\sim\mu/p$ (or, equivalently, $\sim\kappa/p$), and if
$p$ is not too small, both the stress $\BS S$ and heat flux $\BS
q$ are very rapidly damped by their collision operators towards their
Maxwellian values (that is, zeros). This allows to simplify the
transport equations for the stress \eqref{eq:stress_equation} and heat
flux \eqref{eq:heat_flux_equation} into
\begin{subequations}
\label{eq:simplified_stress_heat_flux}
\begin{equation}
\parderiv{\BS S}t+2p\symtr{\nabla\otimes\BS u}\approx -\frac p\mu
\BS S,
\end{equation}
\begin{equation}
\parderiv{\BS q}t+\frac 52 p\nabla \left(\frac p\rho\right)\approx
-\frac 52\frac p\kappa\BS q,
\end{equation}
\end{subequations}
where the advection terms with $\BS S$, $\BS q$ and $\BS Q$ were
dropped, and the contracted fourth-order moment $\BS R$ was replaced
with its Maxwellian value
\begin{equation}
\BS R\approx\frac 52\frac{p^2}\rho\BS I.
\end{equation}
Now, assuming that $\mu$ and $\kappa$ are small enough so that
$\BS S$ and $\BS q$ decay on a much shorter time scale than the
evolution of $\rho$, $\BS u$ and $p$, one then approximates the stress
and heat flux by their approximate steady states:
\begin{subequations}
\label{eq:Navier_Stokes_stress_heat_flux}
\begin{equation}
\label{eq:Navier_Stokes_stress}
\BS S=-2\mu\symtr{\nabla\otimes\BS u},
\end{equation}
\begin{equation}
\BS q=-\kappa\nabla(p/\rho).
\end{equation}
\end{subequations}
A similar procedure is described by Grad \cite{Gra}, pp. 371--372, and
by Struchtrup and Torrilhon \cite{StruTor}.

The important consequence here is that the attracting states for
$\BS S$ and $\BS q$ are not zeros (as opposed to the Euler
closure) due to the effect of advection, which was taken into
account. Lastly, substitution of the approximate states in
\eqref{eq:Navier_Stokes_stress_heat_flux} into the transport equations
\eqref{eq:density_equation}--\eqref{eq:pressure_equation} yields the
famous Navier-Stokes equations:
\begin{subequations}
\label{eq:Navier_Stokes_equations}
\begin{equation}
\parderiv\rho t+\div(\rho\BS u)=0,
\end{equation}
\begin{equation}
\parderiv{(\rho\BS u)}t+\div(\rho\BS u\otimes\BS u)+\nabla p=2\,\div
\left(\mu\symtr{\nabla\otimes\BS u}\right),
\end{equation}
\begin{equation}
\parderiv pt+\div(p\BS u)+\frac 23p\,\div\BS u=\frac
43\mu\left\|\symtr{\nabla\otimes\BS u}\right\|^2+\frac
23\div\left(\kappa\nabla(p/\rho)\right).
\end{equation}
\end{subequations}
Observe that now the Laplace diffusion is present in both the momentum
and pressure equations, on the characteristic time scale of
$\sim\mu^{-1}$ (or, equivalently, $\sim\kappa^{-1}$). This Laplace
dissipation has the effect of ``smoothing out'' the velocity and
pressure fields, which tends to slowly increase the global entropy
$S_g[f]$ (or, alternatively, decrease the global relative entropy
$H_g[f]$), making the process more consistent with the findings of
Desvillettes and Villani \cite{DesVil}. The term with $\mu$ in the
pressure equation is ``ill-posed'' (in the sense that it tends to
increase the pressure, without any counter-balancing), and thus is
often dropped.

In what follows, we assume that the Navier-Stokes relations
\eqref{eq:Navier_Stokes_stress_heat_flux} are valid approximations for
the fine-grained stress $\BS S$ and heat flux $\BS q$ at all
times.

\subsection{The Grad closure}

Under certain conditions, the pressure $p$ becomes so small that the
characteristic time scales $\sim\mu/p$ and $\sim\mu^{-1}$ become
comparable. This happens when the fluid is ``rarefied'', that is, the
fluid particles are so spread out in space that the collisions rarely
happen \cite{Cer3}. In this situation, the local entropy $S_l[f]$ and
the global entropy $S_g[f]$ may exhibit comparable time scales of
evolution, and the Navier-Stokes (let alone Euler's) closure becomes
inapplicable. In particular, one can no longer assume that the
solution $f$ of the Boltzmann equation in
\eqref{eq:boltzmann_equation} is a local Maxwellian
\eqref{eq:Maxwellian}, or anywhere near it.

An appropriate closure for this situation was suggested by Grad
\cite{Gra,Gra2}. The Grad approximation is based on the Hilbert
expansion of the distribution function $f$ in the Boltzmann equation
around its local Maxwellian state \eqref{eq:Maxwellian}, such
that it has the prescribed stress and heat flux, in addition to the
density, momentum and pressure. The resulting distribution
approximates the statistical state of the fluid away from the local
thermodynamic equilibrium and is given by
\begin{equation}
\label{eq:grad_state}
f_G(\BS v)=f_M(\BS v)\left[1+\frac{ \rho}{p^2}\left(\frac{\|\BS v-\BS
    u\|^2}{5p/\rho}-1 \right)\BS q\cdot(\BS v-\BS u)+\frac\rho{2p^2}
  \BS S:((\BS v-\BS u)\otimes(\BS v-\BS u))\right],
\end{equation}
where $f_M$ is the corresponding Maxwellian distribution
\eqref{eq:Maxwellian}. As a result, the full skewness tensor $\BS Q$
and the contracted flatness matrix $\BS R$ are approximated by their
values provided by \eqref{eq:grad_state}, which can be computed
explicitly as:
\begin{subequations}
\begin{equation}
(\BS Q)_{ijk}=\frac 25\left((\BS q)_i\delta_{jk}+(\BS q)_j
  \delta_{ik}+(\BS q)_k\delta_{ij}\right),
\end{equation}
\begin{equation}
\BS R=\left(\frac {5p^2}{2\rho}\BS I+\frac{7p}{2\rho}
\BS S\right).
\end{equation}
\end{subequations}
In particular, this means
\begin{subequations}
\begin{equation}
\div\BS Q=\frac 45\symtr{\nabla\otimes\BS q}+\frac 23(\div \BS q)\BS
I,
\end{equation}
\begin{equation}
\BS Q:(\nabla\otimes\BS u)=\frac 25\left((\div\BS u)\BS q+ (\BS
q\cdot\nabla)\BS u+(\nabla\otimes\BS u)\BS q\right).
\end{equation}
\end{subequations}
Substituting these Grad approximations into the transport equations
\eqref{eq:transport_equations}, we obtain the full set of closed
transport equations for the coarse-grained variables $\rho$, $\BS u$,
$p$, $\BS S$, $\BS q$:
\begin{subequations}
\label{eq:Grad_equations}
\begin{equation}
\parderiv\rho t+\div(\rho\BS u)=0,
\end{equation}
\begin{equation}
\parderiv{(\rho\BS u)}t+\div(\rho\BS u\otimes\BS u+p\BS I+\BS S)
=\BS 0,
\end{equation}
\begin{equation}
\parderiv pt+\div(p\BS u)+\frac 23\left(p\,\div\BS u+\BS S
  :(\nabla \otimes\BS u)+\div\BS q\right)=0,
\end{equation}
\begin{equation}
\parderiv{\BS S}t+(\BS u\cdot\nabla)\BS S +\div(\BS u)
\BS S+2\symtr{\BS S(\nabla \otimes\BS u)} +\frac 45
\symtr{\nabla\otimes\BS q}+2p\symtr{\nabla\otimes\BS u}= \BS C_S,
\end{equation}
\begin{multline}
\parderiv{\BS q}t+\frac 75\big(\div(\BS q\otimes\BS u)+(\BS q\cdot
\nabla)\BS u\big)+\frac 25\big((\nabla\otimes\BS u)\BS q-(\BS u\cdot
\nabla)\BS q\big)-\\-\frac 1\rho\left(\frac 52 p\BS I+\BS S
\right)\div(p\BS I+\BS S)+\div\left( \frac {5p^2}{2\rho}\BS I+
\frac{7p}{2\rho} \BS S\right) =\BS c_q.
\end{multline}
\end{subequations} 
Observe that the collision terms $\BS C_S$ and $\BS c_q$ retain
their general form, while before in the Navier-Stokes closure they
were approximated by the linear damping terms. Although the equations
in \eqref{eq:Grad_equations} are known to develop shock waves at high
Mach numbers \cite{Gra,Gra2}, a suitable regularization have already
been developed by Struchtrup and Torrilhon
\cite{Stru,StruTor,TorStru}, by applying a similar type of closure as
in
\eqref{eq:simplified_stress_heat_flux}--\eqref{eq:Navier_Stokes_stress_heat_flux},
but to the higher-order transport equations for $\BS Q$ and $\BS R$,
which produced additional diffusion terms in the equations for the
stress and heat flux.

\subsection{Higher-order moment closures}

Obviously, one does not have to close the hierarchy in
\eqref{eq:transport_equations} at the equations for the stress
$\BS S$ and heat flux $\BS q$, and instead choose to include the
transport equations for $\BS Q$, $\BS R$ and higher-order moments
\cite{MulRug2,JouCasLeb,AuTorWei,SeeHof,RahStru}. However, for the
clarity of presentation, we restrict ourselves here to the relatively
simple closures presented above.

\section{The moments of the Reynolds-averaged Boltzmann equation}
\label{sec:coarse_grained}

In a turbulent flow, the quantities $\rho$, $\BS u$, $p$, etc usually
exhibit rapid small-scale oscillations in both space and time (that
is, they are ``noisy''), which often renders the direct numerical
simulation of such flow computationally intractable, due to the need
of extremely fine space and time discretization. Instead, one resorts
to solving a {\em different system of transport equations}, which
resolves the large scale features of the flow while forgoing the small
scale oscillations in favor of a more smooth (and, of course,
approximate) solution, in order to afford a more coarse computational
discretization. Below, we refer to such approximate ``smoothed out''
variables as the {\em coarse-grained} variables, superscripted by
asterisks.

Historically, such coarse-grained flow equations where obtained by
applying the operation of ``averaging'' onto the turbulent
Navier-Stokes equations, which was purported to smooth out small scale
rapid oscillations. Reynolds \cite{Rey} pioneered the use of such
averaging operators back in 1895, and they became therefore named
after him.  Throughout the twentieth century, multiple different
interpretations of the Reynolds averaging operator were developed,
including time averaging for steady coarse-grained flows, running
time-average for unsteady flows, scale-dependent spatial averaging
(which later evolved into the large eddy simulation method \cite{Sag},
or LES), and so forth. Eventually, the idea of statistical Reynolds
averaging was adopted (see \cite{MonYag,MonYag2} and references
therein), due to its convenient mathematical properties.

In the current work, we formally assume that the Reynolds averaging
operator is a linear operator, conventionally denoted by an overbar,
with the following properties:
\begin{enumerate}
\item Linearity: for two variables $h_1$ and $h_2$, and two numbers
  $a_1$ and $a_2$,
\begin{equation}
\overline{a_1h_1+a_2h_2} =a_1\overline{h_1}+a_2\overline{h_2}.
\end{equation}
Additionally, $\overline a=a$ if $a$ is a number.
\item Commutativity with respect to differentiation and integration:
\begin{equation}
\overline{\partial h}=\partial\overline h,\qquad\overline{\int
  h}=\int\overline h.
\end{equation}
\item The Reynolds property:
\begin{equation}
\overline{\overline{h_1}h_2}=\overline{h_1}\,\overline{h_2}.
\end{equation}
\end{enumerate}
We also formally assume that the Reynolds-averaged quantities
$\overline\rho$, $\overline{\BS u}$, $\overline p$, as well as their
Reynolds-averaged products (such as $\overline{p\BS u}$, for example)
are sufficiently smooth at small scales to be computationally
tractable. However, due to inherent nonlinearity in the velocity
variable $\BS u$, the transport equations for these Reynolds-averaged
variables involve new unresolved variables which are the Reynolds
averages of products of turbulent fluctuations between the
fine-grained and averaged variables. For example, the averaged
momentum equation involves the Reynolds-averaged outer product of the
turbulent velocity fluctuations, called the ``Reynolds stress'', for
which a closure is usually sought in the form of the Boussinesq
approximation \cite{Bou}. In this work we present a different
approach, which produces the equations for the coarse-grained
variables with a better closure potential.

Here we formally apply the Reynolds operator, described above, to the
Boltzmann equation in \eqref{eq:boltzmann_equation} (rather than the
Navier-Stokes equations), as was done previously by Girimaji
\cite{Gir}. The main novelty of our work is that, unlike \cite{Gir},
we further rebuild the hierarchy of the transport equations like the
one in \eqref{eq:transport_equations} for the moments of the
Reynolds-averaged Boltzmann distribution. The resulting coarse-grained
moment variables are different from the usual Reynolds averages of
conventional moments in \eqref{eq:moments}, and we provide suitable
formulas to relate the new coarse-grained moment variables to the
Reynolds averages of the conventional moments. The key advantage of
our approach is that the transport equations for the new
coarse-grained moments are identical to the conventional moment
transport equations in \eqref{eq:transport_equations} (collision terms
being an exception), and, therefore, naturally lend themselves to the
variety of previously developed closures for
\eqref{eq:transport_equations}. The only unspecified closure
parameters originate from the nonlinear collision term of the
Boltzmann equation, which is not explicitly tractable by a linear
averaging operator.

We define the new coarse-grained variables $\rho^*$, $\BS u^*$, $p^*$,
$\BS S^*$, $\BS q^*$, $\BS Q^*$, $\BS R^*$ as follows:
\begin{subequations}
\begin{equation}
\rho^*=\langle 1\rangle_{\overline f},\qquad\text{coarse-grained density},
\end{equation}
\begin{equation}
\rho^*\BS u^*=\langle\BS v\rangle_{\overline f},\qquad
\text{coarse-grained momentum},
\end{equation}
\begin{equation}
p^*=\frac 13\langle\|\BS v-\BS u^*\|^2\rangle_{\overline f},\qquad
\text{coarse-grained pressure},
\end{equation}
\begin{equation}
\BS S^*=\langle(\BS v-\BS u^*)\otimes(\BS v-\BS u^*)
\rangle_{\overline f}-p^*\BS I, \qquad\text{coarse-grained stress},
\end{equation}
\begin{equation}
\BS q^*=\frac 12\langle\|\BS v-\BS u^*\|^2(\BS v-\BS
u^*)\rangle_{\overline f}, \qquad\text{coarse-grained heat flux},
\end{equation}
\begin{equation}
\BS Q^*=\langle(\BS v-\BS u^*)\otimes(\BS v-\BS u^*)\otimes(\BS v-\BS u^*)
\rangle_{\overline f}, \qquad\text{coarse-grained skewness},
\end{equation}
\begin{equation}
\BS R^*=\frac 12\langle\|\BS v-\BS u^*\|^2(\BS v-\BS u^*)\otimes(\BS v-\BS
u^*)\rangle_{\overline f}, \qquad\text{coarse-grained contracted flatness}.
\end{equation}
\end{subequations}
Then, the following straightforward result follows.
\begin{proposition}[Coarse-grained transport equations]
The new coarse-grained variables are related to the Reynolds averages
of the fine-grained variables as follows:
\begin{subequations}
\label{eq:coarse_grained_variables}
\begin{equation}
\rho^*=\overline{\rho},
\end{equation}
\begin{equation}
\rho^*\BS u^*=\overline{\rho\BS u},\qquad\BS u'=\BS u-\BS
u^*,
\end{equation}
\begin{equation}
\label{eq:cg_p}
p^*=\overline p+\frac 13\overline{\rho\|\BS u'\|^2},
\end{equation}
\begin{equation}
\BS S^*=\overline{\BS S}+\overline{\rho\BS u'\otimes\BS u'}
-\frac 13\overline{\rho\|\BS u'\|^2}\BS I,
\end{equation}
\begin{equation}
\BS q^*=\overline{\BS q}+\overline{\BS S\BS u'}+\frac 52
\overline{p\BS u'}+\frac 12\overline{\rho\|\BS u'\|^2\BS u'},
\end{equation}
\begin{equation}
\BS Q^*=\overline{\BS Q}+\overline{(p\BS I+\BS S)\otimes\BS u'}
+\overline{(p\BS I+\BS S)\otimes\BS u'}^T+\overline{(p\BS I+
  \BS S)\otimes\BS u'}^{TT}+\overline{\rho\BS u'\otimes\BS u'
  \otimes\BS u'},
\end{equation}
\begin{multline}
\BS R^*=\overline{\BS R}+\overline{\BS Q\BS u'}+\overline{(\BS S
  \BS u'+\BS q)\otimes\BS u'}+\overline{\BS u'\otimes(\BS S\BS u'
  +\BS q)}+\\+\frac 72\overline{p\BS u'\otimes\BS u'}+\frac 12
\overline{\|\BS u'\|^2(p\BS I+\BS S)}+\frac 12\overline{\rho\|\BS
  u'\|^2\BS u'\otimes\BS u'}.
\end{multline}
\end{subequations}
Above, $\BS A^T$ and $\BS A^{TT}$ denote two possible transpositions
of a 3-rank tensor $\BS A$. The transport equations for the
coarse-grained variables in \eqref{eq:coarse_grained_variables} are of
the same form as the moment transport equations in
\eqref{eq:transport_equations}, namely
\begin{subequations}
\label{eq:coarse_grained_transport_equations}
\begin{equation}
\label{eq:cg_density}
\parderiv{\rho^*}t+\div(\rho^*\BS u^*)=0,
\end{equation}
\begin{equation}
\parderiv{(\rho^*\BS u^*)}t+\div(\rho^*\BS u^*\otimes\BS u^*+p^*\BS I
+\BS S^*)=\BS 0,
\end{equation}
\begin{equation}
\label{eq:cg_pressure}
\parderiv{p^*}t+\div(p^*\BS u^*)+\frac 23\left[p^*\,\div\BS u^*
  +\BS S^*:(\nabla \otimes\BS u^*)+\div\BS q^*\right]=0,
\end{equation}
\begin{multline}
\label{eq:cg_stress}
\parderiv{\BS S^*}t+(\BS u^*\cdot\nabla)\BS S^* +\div(\BS
u^*)\BS S^*+2\symtr{\BS S^*(\nabla \otimes\BS u^*)}
+\\+\div\BS Q^*-\frac 23(\div\BS q^*)\BS I+2p^*\symtr{\nabla\otimes\BS
  u^*}=\BS C^*_S,
\end{multline}
\begin{multline}
\label{eq:cg_heat_flux}
\parderiv{\BS q^*}t+\div(\BS q^*\otimes\BS u^*)+(\BS q^*\cdot\nabla)
\BS u^*-\frac 1{\rho^*}\left(\frac 52 p^*\BS I+\BS S^*\right)\div
(p^* \BS I+\BS S^*)+\\+\BS Q^*:(\nabla\otimes\BS u^*)+\div\BS R^*
=\BS c^*_q,
\end{multline}
\end{subequations}
where the coarse-grained collision terms are given by the Reynolds
averages
\begin{equation}
\label{eq:coarse_grained_collision}
\BS C^*_S=\overline{\BS C_S},\qquad\BS c^*_q=
\overline{\BS c_q+\BS C_S\BS u'}.
\end{equation}
\begin{proof}
The transport part of \eqref{eq:coarse_grained_transport_equations} is
identical to that of \eqref{eq:transport_equations} since the
transport parts of the underlying Boltzmann's equations for $f$ and
$\overline f$ are identical. To derive the formulas in
\eqref{eq:coarse_grained_variables}, first observe the following
relations:
\begin{subequations}
\begin{equation}
\rho^*=\langle 1\rangle_{\overline f}=\overline{\langle
  1\rangle_f}=\overline\rho,
\end{equation}
\begin{equation}
\rho^*u^*=\langle\BS v\rangle_{\overline f}=\overline{\langle\BS v
  \rangle_f}=\overline{\rho\BS u},
\end{equation}
\begin{equation}
p^*+\frac 13\rho^*\|\BS u^*\|^2=\frac 13\langle\|\BS v\|^2
\rangle_{\overline f}=\frac 13\overline{\langle\|\BS
  v\|^2\rangle_f}=\overline p+\frac 13\overline{\rho\|\BS u\|^2},
\end{equation}
\begin{equation}
\BS S^*+p^*\BS I+\rho^*\BS u^*\otimes\BS u^*=\langle\BS v\otimes
\BS v\rangle_{\overline f}=\overline{\langle\BS v\otimes\BS v
  \rangle_f}=\overline{\BS S}+\overline p\BS I+\overline{\rho\BS
  u\otimes\BS u},
\end{equation}
\begin{multline}
\BS q^*+\BS S^*\BS u^*+\frac 52p^*\BS u^*+\frac 12\rho^*\|\BS u^*
\|^2\BS u^*=\frac 12\langle\|\BS v\|^2\BS v\rangle_{\overline f}=\\=
\frac 12\overline{\langle\|\BS v\|^2\BS v\rangle_f}=\overline{\BS q}
+\overline{\BS S\BS u}+\frac 52\overline{p\BS u}+\frac 12
\overline{\rho\|\BS u\|^2\BS u},
\end{multline}
\begin{multline}
\BS Q^*+(p^*\BS I+\BS S^*)\otimes\BS u^*+((p^*\BS I+\BS S^*)
\otimes\BS u^*)^T+ ((p^*\BS I+\BS S^*)\otimes\BS u^*)^{TT}+\\+
\rho^*\BS u^*\otimes\BS u^*\otimes\BS u^* =\langle\BS v\otimes\BS v
\otimes\BS v \rangle_{\overline f}=\overline{\langle\BS v\otimes\BS v
  \otimes\BS v\rangle_f}=\overline{\BS Q}+\\+\overline{(p\BS I+
  \BS S)\otimes\BS u}+ (\overline{(p\BS I+\BS S)\otimes\BS u}
)^T+ (\overline{(p\BS I+\BS S )\otimes\BS u} )^{TT}+\overline{
  \rho\BS u\otimes\BS u\otimes\BS u},
\end{multline}
\begin{multline}
\BS R^*+\BS Q^*\BS u^*+(\BS S^*\BS u^*+\BS q^*)\otimes\BS u^*+\BS
u^*\otimes(\BS S^*\BS u^*+\BS q^*)+\frac 72p^*\BS u^*\otimes\BS
u^*+\\+\frac 12\|\BS u^*\|^2(p^*\BS I+ \BS S^*)+\frac
12\rho^*\|\BS u^*\|^2\BS u^*\otimes\BS u^*=\frac 12\langle\|\BS
v\|^2\BS v\otimes\BS v \rangle_{\overline f}=\frac
12\overline{\langle\|\BS v\|^2\BS v \otimes\BS v \rangle_f}=\\=
\overline{\BS R}+\overline{\BS Q\BS u}+\overline{(\BS S\BS u+\BS
  q)\otimes\BS u}+\overline{\BS u\otimes(\BS S\BS u+\BS q)}+\frac
72\overline{p\BS u\otimes\BS u}+\frac 12\overline{\|\BS u\|^2(p\BS
  I+\BS S)} +\\+\frac 12\overline{\rho\|\BS u\|^2\BS u\otimes\BS
  u}.
\end{multline}
\end{subequations}
Rearranging the above relations and noting that $\overline{\rho\BS
  u'}=\BS 0$, one obtains the definitions for the coarse-grained
variables in \eqref{eq:coarse_grained_variables}. Lastly, let us
derive the formulas for the coarse-grained collision terms. The
Reynolds-averaged moment transport equation in
\eqref{eq:moment_transport_equation} is given by
\begin{equation}
\parderiv{}t\langle b\rangle_{\overline f}+\div_{\BS x}\langle b\BS
v\rangle_{\overline f}=\langle b \rangle_{\overline{\coll(f)}}.
\end{equation}
In particular, for the stress matrix it is given by
\begin{equation}
\parderiv{}t\langle \BS v\otimes\BS v\rangle_{\overline f}+\div_{\BS
  x}\langle \BS v\otimes\BS v\otimes \BS v\rangle_{\overline
  f}=\overline{\BS C_S},
\end{equation}
where the collision term is unchanged, since the combinations of
equations, added to the stress equation, have no collision terms. Now,
subtracting the same combinations of the transport equations for the
coarse-grained variables (which also do not have collision terms) in
reverse order, we obtain the coarse-grained stress equation with the
same averaged collision term. For the heat flux equation the situation
is slightly different, as it is given by
\begin{equation}
\frac 12\parderiv{}t\langle\|\BS v\|^2\BS v\rangle_{\overline f}+\frac
12\div_{\BS x}\langle\|\BS v\|^2\BS v\otimes\BS v\rangle_{\overline f}
=\overline{\BS c_q+\BS C_S\BS u},
\end{equation}
as the derivation of the moment equation above requires adding the
stress equation, multiplied by $\BS u$, to the heat flux
equation. Now, the derivation of the coarse-grained heat flux equation
from the equation above requires subtracting the coarse-grained stress
equation, accordingly multiplied by the coarse-grained velocity $\BS
u^*$. This results in the coarse-grained collision terms of the form
\eqref{eq:coarse_grained_collision}.
\end{proof}
\end{proposition}
Unfortunately, there appears to be little we can do about the
Reynolds-averaged collision terms in the context of turbulence
transport. While the transport part of the Boltzmann equation is
linear and thus is ``transparent'' to the Reynolds averaging, the same
is not true for the collision terms. Without more information about
the properties of the statistical ensemble (and, therefore,
statistical physics of the fluid under consideration), we can only
conclude that the coarse-grained collision terms $\BS C^*_S$ and
$\BS c^*_q$ have a general dissipative effect towards the global
maximum entropy state under the total mass, momentum and energy
constraints. Below in Section \ref{sec:computational_considerations}
we suggest crude parameterizations of $\BS C^*_S$ and $\BS c^*_q$
based on dimensional arguments.

\section{The transport of turbulent energy under the anelastic
turbulence approximation}
\label{sec:turbulent_fluctuations}

In the previous section we derived a hierarchy of the coarse-grained
equations \eqref{eq:coarse_grained_transport_equations}, which does
not seem to require anything other than a suitable closure to become
computationaly tractable. However, observe that the new coarse-grained
variables in \eqref{eq:coarse_grained_variables} are not exactly what
one would naturally need to model. For example, a meteorologist would
likely need to model, among other variables, the Reynolds average of
the pressure $\overline p$ (as the closest computable alternative to
the pressure $p$ itself), yet observe that the coarse-grained variable
$p^*$ is not $\overline p$, but rather $\overline
p+\overline{\rho\|\BS u'\|^2}/3$ (see \eqref{eq:cg_p}), where the
additional term is not zero!

Therefore, we need to provide a means to recover the Reynolds averages
of the fine-grained variables from the computed coarse-grained
variables. In order to do this, we make an additional simplifying
approximation about the nature of the small-scale turbulent
fluctuations, which absolves us of the need to address the Reynolds
averages of the density and velocity variables, leaving only the
pressure, stress and heat flux to work with. Above, in the derivation
of the equations for the coarse-grained variables, we assumed the most
general form of turbulence, were the turbulent fluctuations were
present in all fluid variables, and thus the Reynolds averages of
fluid variables were generally never identity operators in the
presence of turbulence. However, in many applications it is known that
the density $\rho$ is weakly affected by the small-scale turbulent
fluctuations, and varies only on the coarse-grained ``resolved''
scale, while the higher order moments, such as the velocity $\BS u$,
pressure $p$, etc are still affected by the turbulence. Thus, here we
assume the identity $\rho=\rho^*$, and, therefore, $\rho$ can be
factored out of the Reynolds averages. As a result, the turbulent
velocity component $\BS u'=\BS u-\BS u^*$ becomes anelastic:
\begin{equation}
\label{eq:anelastic_approximation}
\div(\rho\BS u')=0.
\end{equation}
Indeed, observe that subtracting the coarse-grained density equation
\eqref{eq:cg_density} from the fine-grained density equation
\eqref{eq:density_equation} results in
\begin{equation}
\parderiv{}t(\rho-\rho^*)=\div(\rho\BS u-\rho^*\BS u^*).
\end{equation}
Setting $\rho^*=\rho$ above yields \eqref{eq:anelastic_approximation}.

The anelastic turbulence approximation does not change the form of the
coarse-grained transport equations in
\eqref{eq:coarse_grained_transport_equations}, but instead simplifies
the definitions of the coarse-grained variables in
\eqref{eq:coarse_grained_variables} to some extent. For the Reynolds
averages of the anelastic turbulent fluctuations
\begin{equation}
\BS u'=\BS u-\BS u^*,\qquad p'=p-p^*,\qquad\BS S'=\BS S
-\BS S^*,\qquad\BS q'=\BS q-\BS q^*,
\end{equation}
we obtain directly
\begin{subequations}
\begin{equation}
\overline{\BS u'}=\BS 0,
\end{equation}
\begin{equation}
\label{eq:pressure_fluctuation}
\overline{p'}=-\frac 13\rho\overline{\|\BS u'\|^2},
\end{equation}
\begin{equation}
\label{eq:stress_fluctuation}
\overline{\BS S'}=-\rho\left(\overline{\BS u'\otimes\BS u'}-\frac
13\overline{\|\BS u'\|^2}\BS I\right),
\end{equation}
\begin{equation}
\label{eq:heat_flux_fluctuation_2}
\overline{\BS q'}=-\left(\overline{\BS S'\BS u'}+\frac 52
\overline{p'\BS u'}+\frac 12\rho\overline{\|\BS u'\|^2\BS u'}\right).
\end{equation}
\end{subequations}
Obviously, computing $\overline{p'}$, $\overline{\BS S'}$ and
$\overline{\BS q'}$ in addition to the coarse-grained variables in
\eqref{eq:coarse_grained_variables} allows to connect back to the
Reynolds averages of the original variables. Here we are going to
start with $\overline{p'}$, which requires its own separate equation.

To derive the transport equation for the Reynolds-averaged turbulent
pressure $\overline{p'}$, we subtract the coarse-grained pressure
equation \eqref{eq:cg_pressure} from the fine-grained pressure
equation \eqref{eq:pressure_equation} and obtain
\begin{multline}
\parderiv{p'}t+\div\left(p^*\BS u'+p'\BS u^*+p'\BS u'\right)+\frac 23
\left(p^*\div\BS u'+p'\div\BS u^*+p'\div\BS u'\right)+\\+\frac 23
\left(\BS S:(\nabla\otimes\BS u)-\BS S^*:(\nabla\otimes\BS
u^*)\right)+\frac 23\div\BS q'=0.
\end{multline}
Applying the Reynolds average to the equation above results in
\begin{multline}
\parderiv{\overline{p'}}t+\div\left(\overline{p'}\BS u^*\right)+\frac
23\overline{p'}\div\BS u^*+\div\left(\overline{p'\BS u'}\right)+\frac
23\overline{p'\div\BS u'}+\\+\frac 23\left(\overline{\BS S:(\nabla
  \otimes \BS u)}-\BS S^*:(\nabla\otimes\BS u^*)\right)+\frac 23
\div\overline{\BS q'}=0.
\end{multline}
We now observe that
\begin{subequations}
\begin{equation}
p'\div\BS u'=\frac{p'}\rho\rho\,\div\BS u'=\frac{p'}\rho(\div(\rho\BS
u')-\BS u'\cdot\nabla\rho)=-p'\BS u'\cdot\frac{\nabla\rho}\rho,
\end{equation}
\begin{equation}
\overline{\BS S:(\nabla\otimes\BS u)}-\BS S^*:(\nabla\otimes
\BS u^*)=\overline{\BS S'}:(\nabla\otimes\BS u^*)+
\overline{\BS S':(\nabla\otimes\BS u')},
\end{equation}
\end{subequations}
which further results in
\begin{multline}
\parderiv{\overline{p'}}t+\div\left(\overline{p'}\BS u^*\right)+\frac
23 \overline{p'}\div\BS u^*+\div\left(\overline{p'\BS u'}\right)-\frac
23\frac{\nabla\rho}\rho\cdot\overline{p'\BS u'}+\\+\frac 23 \left(
\overline{\BS S'}:(\nabla\otimes\BS u^*)+\overline{\BS S'
  :(\nabla \otimes\BS u')}+\div\overline{\BS q'}\right)=0.
\end{multline}
The turbulent pressure $\overline{p'}$ is not a convenient measure of
turbulence from a physicist's perspective, since, according to
\eqref{eq:pressure_fluctuation}, it cannot be positive. Thus, it is
more convenient to replace it with a nonnegative quantity, and here we
choose the kinetic energy of turbulent velocity fluctuations
\cite{MonYag,MonYag2} for this purpose:
\begin{equation}
\label{eq:turbulent_energy}
E_t=-\frac 32\overline{p'}=\frac 12\rho\overline{\|\BS u'\|^2}.
\end{equation}
For the turbulent energy in \eqref{eq:turbulent_energy}, the transport
equation becomes
\begin{multline}
\parderiv{E_t}t+\div(E_t\BS u^*)+\frac 23 E_t\div\BS u^*-\frac 32\div(
\overline{p'\BS u'})+\frac{\nabla\rho}\rho\cdot\overline{p'\BS u'}-
\\- \overline{ \BS S'}:(\nabla\otimes\BS u^*)-\overline{
  \BS S':(\nabla \otimes\BS u')}-\div\overline{\BS q'}=0.
\end{multline}
The term $\overline{\BS S':(\nabla\otimes\BS u')}$ in the equation
above can be rearranged as follows. We assume that the microscale
viscosity $\mu$ can be factored out of the Reynolds average (that is,
replacing $\mu$ with its own Reynolds average is an acceptable
simplification). Then, one writes
\begin{multline}
\label{eq:turb_sigma_u}
\overline{\BS S':(\nabla\otimes\BS u')}=\overline{\BS S:(
  \nabla \otimes\BS u')}=-2\mu\overline{\symtr{\nabla \otimes\BS u}
  :(\nabla\otimes\BS u')}=\\=-2\mu\overline{\symtr{\nabla \otimes \BS
    u'}:(\nabla\otimes\BS u')}=\frac 23\mu \overline{(\div
  \BS u')^2}-\mu\overline{(\nabla\otimes\BS u') :(\nabla
  \otimes\BS u')^T}-\mu\overline{\left\|\nabla \otimes\BS
  u'\right\|^2},
\end{multline}
where we replaced the fine-grained stress $\BS S$ with its
Navier-Stokes approximation \eqref{eq:Navier_Stokes_stress}. For the
first term in the right-hand side of \eqref{eq:turb_sigma_u} we use
the anelastic turbulence approximation \eqref{eq:anelastic_approximation}
and obtain
\begin{multline}
\overline{(\div\BS u')^2}=\frac 1{\rho^2}\overline{(\rho\div\BS u')^2}
=\frac 1{\rho^2} \overline{(\div(\rho\BS u')-\nabla\rho\cdot\BS u')^2}
=\frac 1{\rho^2}(\nabla\rho \otimes\nabla\rho):\overline{(\BS u'
  \otimes\BS u')}=\\=-(\overline{\BS S'}+\overline{p'}\BS I):\frac
1 {\rho^3}(\nabla\rho \otimes\nabla\rho)=\frac 23\frac{\|\nabla\rho
  \|^2}{\rho^3}E_t-\frac 1{\rho^3}(\nabla\rho\otimes\nabla\rho):
\overline{\BS S'},
\end{multline}
where we used \eqref{eq:stress_fluctuation} and
\eqref{eq:turbulent_energy} in the last equality. For the second term
in the right-hand side of \eqref{eq:turb_sigma_u} we obtain
\begin{multline}
\overline{(\nabla\otimes\BS u') :(\nabla \otimes\BS u')^T}=(\nabla
\otimes\nabla):\overline{(\BS u'\otimes\BS u')}+2\div\left(\overline{(
  \BS u'\otimes\BS u')}\frac{\nabla\rho}\rho\right)+\\+\frac 1{\rho^2}
(\nabla\rho\otimes\nabla\rho):\overline{(\BS u'\otimes\BS u')} =
(\nabla\otimes\nabla):\left(\frac 2{3\rho} E_t\BS I-\frac 1\rho
\overline{\BS S'}\right)+2\div\left(\frac 2{3\rho^2} E_t\nabla
\rho-\overline{\BS S'}\frac{\nabla\rho}{\rho^2}\right)+\\+\frac
1{\rho^2}(\nabla\rho\otimes\nabla\rho):\left(\frac 2{3\rho} E_t\BS I
-\frac 1\rho\overline{\BS S'}\right)=\frac 23\Delta\left(\frac{
  E_t}\rho\right)+\frac 43\div\left( E_t\frac{\nabla\rho}{\rho^2}
\right)+\frac 23\frac{\|\nabla\rho\|^2}{\rho^3} E_t-\\-(\nabla\otimes
\nabla):\left(\frac{\overline{ \BS S'}}\rho \right)-2\div\left(
\overline{\BS S'}\frac{\nabla\rho}{\rho^2}\right)-\frac 1{\rho^3}
(\nabla\rho\otimes\nabla\rho):\overline{\BS S'}=\frac 23\frac{
  \Delta E_t}\rho+\frac 23 E_t\frac{\Delta\rho}{\rho^2}-\\-\frac 23
E_t \frac{\|\nabla\rho\|^2}{\rho^3}-\frac 1\rho(\nabla\otimes\nabla):
\overline{\BS S'}-\frac 1{\rho^2}\big((\nabla\otimes\nabla)\rho
\big):\overline{\BS S'}+\frac 1{\rho^3}(\nabla\rho\otimes
\nabla\rho):\overline{\BS S'}.
\end{multline}
Now, we denote the last term in the right-hand side of
\eqref{eq:turb_sigma_u} as the turbulent energy dissipation rate
\cite{MonYag,MonYag2,Wilc},
\begin{equation}
\rho\varepsilon_t=\mu\overline{\left\|\nabla\otimes\BS
  u'\right\|^2},
\end{equation}
as it cannot be expressed in terms of the Reynolds averages of the
turbulent fluctuations that we already defined. This leads to
\begin{multline}
\overline{\BS S':(\nabla\otimes\BS u')}=-\frac 23\frac\mu\rho
\Delta E_t+\frac 23\frac\mu\rho\left(\frac 53\frac{\|\nabla\rho\|^2}{
  \rho^2}-\frac{\Delta\rho}\rho\right)E_t+\\+\frac\mu\rho(\nabla\otimes
\nabla):\overline{\BS S'}+\frac\mu{\rho^2}\big((\nabla\otimes
\nabla)\rho\big): \overline{\BS S'}-\frac{5\mu}{3\rho^3}(\nabla
\rho\otimes\nabla\rho):\overline{\BS S'}-\rho\varepsilon_t,
\end{multline}
and, consequently,
\begin{multline}
\label{eq:turbulent_energy_equation_2}
\parderiv{E_t}t+\div(E_t\BS u^*)+\frac 23\left(\div\BS u^*+\frac\mu
\rho\Delta+\mu\frac{\Delta\rho}{\rho^2}-\frac 53\mu\frac{\|\nabla\rho
  \|^2}{\rho^3}\right)E_t-\\-\left(\nabla\otimes\BS u^*+\frac\mu\rho(
\nabla\otimes\nabla)+\frac\mu{\rho^2}\big((\nabla\otimes\nabla)\rho
\big)-\frac{5\mu}{3\rho^3}(\nabla\rho\otimes\nabla\rho)\right):
\overline{\BS  S'}-\\-\div\overline{\BS q'}-\frac 32\div\left(
\overline{p'\BS u'}\right)+\frac{\nabla\rho}\rho\cdot\overline{p'\BS
  u'}+ \rho\varepsilon_t =0.
\end{multline}
Observe that the equation for the turbulent energy above depends on
the Reynolds-averaged turbulent stress $\overline{\BS S'}$, heat
flux $\overline{\BS q'}$, the quantity $\overline{p'\BS u'}$, as well
as the turbulent energy dissipation rate $\rho\varepsilon_t$.

\subsection{Isotropic turbulence assumption and its limitations}

As a special case of the turbulent flow, here we consider a
hypothetical situation where the turbulent velocity fluctuations $\BS
u'$ at each spatial point are statistically decorrelated and
isotropic. This implies the following conditions for statistical
averages of turbulent quantities:
\begin{equation}
\label{eq:isotropic_assumption}
\overline{\BS u'\otimes\BS u'}=\frac 13\|\BS u'\|^2\BS
I,\qquad\overline{\BS S'}=\BS 0,\qquad \overline{p'\BS
  u'}=\overline{\BS q'}=\BS 0.
\end{equation}
The first equality above is due to the fact that the
cross-correlations between different components of the turbulent
velocity are zeros, while self-correlations are equal. The second
equality follows immediately from the first one by using
\eqref{eq:stress_fluctuation}. The third equality is due to the fact
that, because of isotropy, for each instance of $\BS u'$, $p'$ and
$\BS q'$, there must be another with $-\BS u'$, $p'$ and $-\BS q'$,
and with the same statistical weight. This is, of course, an idealized
assumption, which is unlikely to hold exactly in practical
situations. In fact, it is somewhat of a turbulent analog of the
Maxwellian equilibrium condition in \eqref{eq:Maxwellian} for the
fluid particle velocities, assumed without rigorous justification.

Under the isotropic assumption in \eqref{eq:isotropic_assumption}, the
equation for the turbulent energy transport in
\eqref{eq:turbulent_energy_equation_2} becomes
\begin{equation}
\parderiv{E_t}t+\div(E_t\BS u^*)+\frac 23 \left(\div\BS u^*+\frac\mu
\rho\Delta+\mu\frac{\Delta\rho}{\rho^2}-\frac 53 \mu\frac{\|\nabla
  \rho\|^2}{\rho^3}\right)E_t+\rho\varepsilon_t=0.
\end{equation}
There are two problems with this equation. First, the diffusion term
above is clearly ill-posed, which means that the turbulent energy will
be amplified by this term at small scales, rather than damped. The
second problem with the equation above is that it has no means of
producing turbulent energy from zero initial condition. At the same
time, it is very well known that the turbulence appears even if there
was none to begin with, from the interactions of the large-scale
motions with the small-scale viscous dissipation. Thus, the only way
to create such interaction is to couple the higher-order Reynolds
averages in \eqref{eq:turbulent_energy_equation_2} to the
coarse-grained variables, rather than assuming isotropy. In a broad
sense, the turbulence is produced by the anisotropy of the large-scale
flow.

\subsection{Coupling the turbulent Reynolds averages to the
coarse-grained variables}

Here we assume that on the fine-grained scale the fluid satisfies the
Navier-Stokes approximations in
\eqref{eq:Navier_Stokes_stress_heat_flux}. We also assume that
replacing the viscosity $\mu$ and heat conductivity $\kappa$ in
\eqref{eq:Navier_Stokes_stress_heat_flux} with their corresponding
Reynolds averages is an acceptable approximation. Then, the Reynolds
averages of the turbulent fluctuations $\overline{\BS S'}$ and
$\overline{\BS q'}$ can be expressed via the coarse-grained stress
$\BS S^*$ and heat flux $\BS q^*$, and the Reynolds-averaged
Navier-Stokes approximations for the fine-grained stress and heat flux
as follows:
\begin{subequations}
\label{eq:sigma_q_fluctuations}
\begin{equation}
\overline{\BS S'}=\overline{\BS S}-\BS S^*=-\BS
 S^*-2\mu\symtr{\nabla\otimes\BS u^* },
\end{equation}
\begin{equation}
\overline{\BS q'}=\overline{\BS q}-\BS q^*=-\BS q^*-\kappa\nabla
\left(\frac{\overline p}\rho\right)=-\BS q^*-\kappa\nabla\left(
\frac{p^*}\rho\right)+\frac 23\kappa\nabla\left(\frac{E_t}\rho
\right).
\end{equation}
\end{subequations}
Substituting the above expressions into the transport equation for the
turbulent energy yields
\begin{multline}
\parderiv{E_t}t+\div(E_t\BS u^*)+\frac 23 \left(\div\BS u^*+\frac\mu
\rho\Delta+\mu\frac{\Delta\rho}{\rho^2}-\frac 53\mu\frac{\|\nabla\rho
  \|^2}{\rho^3}\right)E_t-\frac 23\div\left(\kappa\nabla\left(\frac{
  E_t}\rho \right)\right)+\\+\left(\nabla\otimes\BS u^*+\frac\mu\rho
(\nabla\otimes\nabla)+\frac\mu{\rho^2}\big((\nabla\otimes\nabla)\rho
\big)-\frac{5\mu}{3\rho^3}(\nabla\rho\otimes\nabla\rho)\right):\left(
\BS  S^*+2\mu\symtr{\nabla\otimes\BS u^*} \right)+\\+\div\BS q^*+
\div\left(\kappa\nabla\left( \frac{p^*}\rho\right)\right)-\frac 32
\div(\overline{p'\BS u'})+\frac{\nabla\rho} \rho\cdot\overline{p'\BS
  u'}+\rho\varepsilon_t =0.
\end{multline}
The unknown term $\overline{p'\BS u'}$ is still present, and needs
either its own separate transport equation (which will, of course,
involve yet higher-order moment combinations of turbulent quantities),
or a suitable closure. For simplicity, here we propose a closure under
the assumption of Bernoulli's principle. For that, first observe the
following relation:
\begin{multline}
\overline{\BS S'\BS u'}=\overline{\BS S\BS u'}=-2\mu
\overline{\symtr{\nabla\otimes\BS u}\BS u'}=-2\mu\overline{
  \symtr{\nabla\otimes\BS u'}\BS u'}=\\=-\mu\overline{\left(
  \nabla\otimes\BS u'+(\nabla\otimes\BS u')^T-\frac 23(\div\BS u')\BS
  I\right)\BS u'},
\end{multline}
where, as before, the Navier-Stokes approximation is used for the
Reynolds average of the fine-grained stress
$\overline{\BS S}$. For the separate terms above we have, with the
help of \eqref{eq:anelastic_approximation},
\begin{subequations}
\begin{equation}
\overline{(\nabla\otimes\BS u')\BS u'}=-\frac 32\nabla\left(\frac{
  \overline{p'}}\rho\right)=\nabla\left(\frac{E_t}\rho\right),
\end{equation}
\begin{equation}
\overline{(\nabla\otimes\BS u')^T\BS u'}=\frac 1\rho\div(\rho
\overline{\BS u'\otimes\BS u'})=-\frac 1\rho\div\overline{\BS S'}
+\frac 23\nabla\left(\frac{E_t}\rho\right)-\frac 23 E_t\frac{\nabla
  \rho}{\rho^2},
\end{equation}
\begin{equation}
\overline{(\div\BS u')\BS u'}=-\overline{(\BS u'\otimes\BS u')}\frac{
  \nabla\rho}\rho=\overline{\BS S'}\frac{\nabla\rho}{\rho^2}-
\frac 23 E_t\frac{\nabla\rho}{\rho^2},
\end{equation}
\end{subequations}
which, when assembled together, lead to the expression of
$\overline{\BS S'\BS u'}$ in terms of the Reynolds average of the
turbulent stress fluctuation $\overline{\BS S'}$ and the turbulent
kinetic energy $E_t$:
\begin{equation}
\label{eq:sigma_u}
\overline{\BS S'\BS
  u'}=\left(\frac\mu\rho\div+\frac{2\mu}{3\rho^2}\nabla\rho\right)
\overline{\BS S'}-\frac 53\mu\nabla \left(\frac{E_t}\rho\right)
-\frac{10}9\mu E_t\frac{\nabla\rho}{\rho^2}.
\end{equation}
Combining \eqref{eq:sigma_u} with \eqref{eq:heat_flux_fluctuation_2}
and \eqref{eq:sigma_q_fluctuations}, we obtain
\begin{multline}
\BS q^*+\kappa\nabla\left(\frac{p^*}\rho\right)-\frac 23\kappa\nabla
\left(\frac{E_t} \rho\right)=-\overline{\BS q'} = -\left(\frac\mu\rho
\div+\frac{2\mu}{3\rho^2}\nabla\rho\right)\left(\BS S^*+2\mu
\symtr{\nabla\otimes\BS u^* }\right)-\\-\frac 53\mu\nabla\left(\frac{
  E_t}\rho\right)-\frac{10\mu}{9\rho^2}E_t\nabla\rho+\frac
52\overline{p'\BS u'}+\frac 12\rho\overline{\|\BS u'\|^2\BS u'},
\end{multline}
or, after rearranging the terms,
\begin{multline}
\frac 52\overline{p'\BS u'}+\frac 12\rho\overline{\|\BS u'\|^2\BS u'}
=\BS q^* +\kappa\nabla\left(\frac{p^*}\rho\right) +\frac{5\mu-2\kappa
}3\nabla\left(\frac{E_t}\rho\right)+\frac{10\mu}{9\rho^2}E_t\nabla\rho+
\\ +\left(\frac\mu\rho\div+\frac{2\mu}{3\rho^2}\nabla\rho\right)
\left(\BS S^*+2\mu\symtr{\nabla\otimes\BS u^*} \right).
\end{multline}
Observe that the Reynolds averages of turbulent fluctuations in the
left-hand side above are expressed entirely in terms of the
coarse-grained variables and the turbulent energy $E_t$ in the the
right-hand side. However, in order to close the turbulent energy
transport equation in terms of the coarse-grained variables and the
turbulent energy itself, one has to surmise a plausible relation
between the Reynolds averages $\overline{p'\BS u'}$ and
$\overline{\|\BS u'\|^2\BS u'}$. There is certainly more than one way
to do that (as an example, one could postulate $\overline{\|\BS
  u'\|^2\BS u'}=\BS 0$, for a simplest closure), and we here choose
what we perceive as the most physically realistic. Namely, we recall
Bernoulli's principle, which states that the change in the local flow
velocity tends to affect the pressure in the flow in a manner as to
preserve the total energy. If we treat the turbulent velocity
fluctuation $\BS u'$ as the primary cause for the turbulent pressure
fluctuation $p'$ at the same location, Bernoulli's principle then
leads to the relation
\begin{equation}
\label{eq:bernoulli}
p'=-\frac 13\rho\|\BS u'\|^2.
\end{equation}
Of course, we understand that in reality the relation in
\eqref{eq:bernoulli} cannot hold at each spatial point exactly, as
Bernoulli's principle is merely an approximation. However, we note
that the exact relation \eqref{eq:pressure_fluctuation} is precisely
the Reynolds-averaged relation \eqref{eq:bernoulli}. So, we surmise
that it is plausible to incorporate an approximate relation in
\eqref{eq:bernoulli} into the relation between $\overline{p'\BS u'}$
and $\overline{\|\BS u'\|^2\BS u'}$ as follows:
\begin{equation}
\label{eq:bernoulli_p_u}
\overline{p'\BS u'}=-\frac 13\rho\overline{\|\BS u'\|^2\BS
  u'},\qquad\text{or}\qquad\frac 52\overline{p'\BS u'}+\frac
12\rho\overline{\|\BS u'\|^2\BS u'}=\overline{p'\BS u'}.
\end{equation}
In particular, as a result of Bernoulli's assumption, the Reynolds
average of the turbulent heat flux becomes the Reynolds average of the
product of the turbulent velocity with the turbulent pressure matrix,
taken with the opposite sign:
\begin{equation}
\overline{\BS q'}=-\overline{\BS P'\BS u'},\qquad \BS P'=p'\BS I+
\BS S'.
\end{equation}
This results in the following expression for $\overline{p'\BS u'}$ in
terms of the coarse-grained variables and the turbulent energy $E_t$:
\begin{multline}
\overline{p'\BS u'}=\BS q^*+\kappa\nabla\left(\frac{p^*}\rho\right)+
\frac{5\mu-2 \kappa}3\nabla\left(\frac{E_t}\rho\right)+\frac{10\mu}{
  9\rho^2}E_t \nabla\rho+\\+\left(\frac\mu\rho\div+\frac{2\mu}{3
  \rho^2}\nabla\rho\right)\left(\BS S^*+2\mu\symtr{\nabla\otimes
  \BS u^* }\right).
\end{multline}
The turbulent energy equation then becomes closed with respect to the
coarse-grained variables and the turbulent energy $E_t$ itself:
\begin{multline}
\parderiv{E_t}t+\div(E_t\BS u^*)+\frac 23 \left(\div\BS u^*+\frac\mu
\rho\Delta+\mu\frac{\Delta\rho}{\rho^2}-\frac 53\mu\frac{\|\nabla\rho
  \|^2}{\rho^3}\right)E_t-\frac 23\div\left(\kappa\nabla\left(\frac{
  E_t}\rho \right)\right)+\\+\left(\frac{\nabla\rho}\rho-\frac 32
\nabla\right)\cdot\left(\frac{5\mu-2 \kappa}3\nabla\left(\frac{E_t}
\rho\right)+\frac{10\mu}{ 9\rho^2}E_t\nabla\rho\right)+\left(\frac{
  \nabla\rho}\rho-\frac 12\nabla\right)\cdot\left(\BS q^*+\kappa
\nabla\left(\frac{p^*}\rho\right)\right)+\\+\left(\nabla\otimes\BS u^*
+\frac\mu\rho (\nabla\otimes\nabla)+\frac\mu{\rho^2}\big((\nabla
\otimes\nabla)\rho\big)-\frac{5\mu}{3\rho^3}(\nabla\rho\otimes\nabla
\rho)\right):\left(\BS S^*+2\mu\symtr{\nabla\otimes\BS u^*}
\right)+\\+\left(\frac{\nabla\rho}\rho-\frac 32\nabla\right)\cdot
\left(\left(\frac\mu\rho\div+\frac{2\mu}{3\rho^2}\nabla\rho\right)
\left(\BS S^*+2\mu\symtr{\nabla\otimes \BS u^* }\right)\right)+
\rho\varepsilon_t=0.
\end{multline}
Rearranging the terms, we further obtain
\begin{multline}
\label{eq:turbulent_energy_equation}
\parderiv{E_t}t+\div\left(E_t\left(\BS u^*+\frac{15\mu-4\kappa}{3
  \rho^2}\nabla\rho+\frac{\nabla(2\kappa-15\mu)}{6\rho}\right)\right)
+\frac 23 E_t\bigg(\div\BS u^*+\frac{15\mu-2\kappa}2\frac{\|\nabla\rho
  \|^2}{\rho^3}+\\ +(\kappa-4\mu)\frac{\Delta\rho}{\rho^2}+\nabla(
\kappa-5\mu)\cdot\frac{\nabla\rho}{\rho^2}+\div\left(\frac{\nabla(2
  \kappa-15\mu)}{4\rho}\right)\bigg)+\frac{2\kappa-11\mu}{6\rho}
\Delta E_t=F^*_t-\rho\varepsilon_t,
\end{multline}
where $F^*_t$ is the coarse-grained forcing:
\begin{multline}
\label{eq:F_t}
F^*_t=\left(\frac 12\nabla-\frac{\nabla\rho}\rho\right)\cdot \left(\BS
q^*+\kappa\nabla\left(\frac{p^*}\rho\right)\right)-\\-\left(\nabla
\otimes\BS u^* +\frac\mu\rho (\nabla\otimes\nabla)+\frac\mu{\rho^2}
\big((\nabla\otimes\nabla)\rho\big)-\frac{5\mu}{3\rho^3}(\nabla\rho
\otimes\nabla \rho)\right):\left(\BS S^*+2\mu\symtr{\nabla\otimes
  \BS u^*}\right)+\\+\left(\frac 32\nabla-\frac{\nabla\rho}\rho\right)
\cdot\left(\left(\frac\mu\rho\div+\frac{2\mu}{3\rho^2}\nabla\rho
\right)\left(\BS S^*+2\mu\symtr{\nabla\otimes \BS u^* }\right)
\right).
\end{multline}
Here observe the following properties:
\begin{itemize}
\item For the Prandtl number of the ideal gas, $\Pran=2/3$, and the
  assumption that the spatial derivatives of $\mu$ and $\kappa$ can be
  neglected, the advection term in
  \eqref{eq:turbulent_energy_equation} is exactly the same as it is
  for the usual pressure variable.
\item The Laplace diffusion term in
  \eqref{eq:turbulent_energy_equation} is well posed as long as
  $\Pran>5/11$, which includes the ideal gas.
\item The term with $\|\nabla\rho\|^2$ in
  \eqref{eq:turbulent_energy_equation} is linear damping as long as
  $\Pran>1/3$, which again includes ideal gas.
\end{itemize}
For a strong turbulence (that is, large coarse-grained stress
$\BS S^*$ and heat flux $\BS q^*$) and small microscale viscosity
$\mu$ and heat conductivity $\kappa$, one likely can drop the terms
scaled by the latter in \eqref{eq:turbulent_energy_equation} and
\eqref{eq:F_t}, except the one for the diffusion, as it could
potentially be the highest-order differential operator, given the lack
of information on $\rho\varepsilon_t$. The resulting simplified
equation is given by
\begin{equation}
\parderiv{E_t}t+\div(E_t\BS u^*)+\frac 23E_t\div\BS u^*+\frac{2\kappa
  -11\mu}{6\rho}\Delta E_t =\left(\frac 12\nabla-\frac{\nabla\rho}
\rho\right)\cdot\BS q^*-\BS S^*:(\nabla\otimes\BS u^*)
-\rho\varepsilon_t.
\end{equation}
Observe that this simplified transport equation for the turbulent
energy is the same as the one for the Euler approximation of the
pressure equation, but with the additional forcing via the
coarse-grained variables $\BS S^*$ and $\BS q^*$, small scale
dissipation $\rho\varepsilon_t$, and a weak diffusion.

At this point, it becomes clear how the turbulence is produced from an
initial condition which belongs to the fully resolved coarse-grained
scale, with zero stress and heat flux. Schematically, this process can
be illustrated as
\begin{equation}
\left.\begin{array}{c@{\quad\longrightarrow\quad}c}
\displaystyle 2p^*\symtr{\nabla\otimes\BS u^* } & \BS S^*
\\ \displaystyle \frac 52p^*\nabla\left(\frac{p^*}\rho\right) & \BS
q^* \end{array}\right\}\longrightarrow E_t,
\end{equation}
where the are two stages:
\begin{enumerate}
\item As the stress and heat flux are zero initially, the
  coarse-grained strain rate term $2p^*\symtr{\nabla\otimes\BS u^*
    }$ and the coarse-grained temperature gradient term
  $(5/2)p^*\nabla(p^*/\rho)$ act as external forcing in the equations
  for the coarse-grained stress \eqref{eq:cg_stress} and heat flux
  \eqref{eq:cg_heat_flux}, respectively, causing deviation from the
  Maxwellian equilibrium on the coarse-grained scale.
\item In turn, the nonzero coarse-grained stress $\BS S^*$ and
  heat flux act $\BS q^*$ act as external forcing in the turbulent
  energy equation \eqref{eq:turbulent_energy_equation}, causing the
  nonzero turbulent energy to appear.
\end{enumerate}
In particular, this means that the coarse-grained collision terms
$\BS C^*_S$ and $\BS c^*_q$ provide weaker damping rate than
their fine-grained counterparts even when there is no turbulence
developed yet.

\section{Collision terms and moment closures for the coarse-grained
transport equations}
\label{sec:coarse_grained_moment_closures}

Thus far, we developed the following ingredients of the coarse-grained
transport framework for a turbulent flow:
\begin{itemize}
\item A hierarchy of the new coarse-grained variables in
  \eqref{eq:coarse_grained_variables} and the corresponding hierarchy
  of the transport equations in
  \eqref{eq:coarse_grained_transport_equations}. Although the
  expressions for the coarse-grained variables in
  \eqref{eq:coarse_grained_variables} and the corresponding equations
  \eqref{eq:coarse_grained_transport_equations} only include the
  moments up to the heat flux, clearly the full set of moments is
  infinite, and requires either infinitely many transport equations,
  or an approximate closure. The derivation was done under the general
  assumption that, microscopically, the flow is described by the
  Boltzmann equation for a monatomic ideal gas
  \eqref{eq:boltzmann_equation}.
\item The relations for the corresponding Reynolds-averaged turbulent
  fluctuations. Those are given in the form of the turbulent energy
  (or, equivalently, turbulent pressure) transport equation in
  \eqref{eq:turbulent_energy_equation}, and the relations for the
  turbulent stress and heat flux in \eqref{eq:sigma_q_fluctuations}.
  This was done under the following assumptions:
  \begin{enumerate}
    \item Validity of the Navier-Stokes approximations for the
      fine-grained stress and heat flux
      \eqref{eq:Navier_Stokes_stress_heat_flux};
    \item Invariance of the density $\rho$ under the Reynolds
      averaging and the resulting anelastic turbulence approximation
      \eqref{eq:anelastic_approximation};
    \item Approximation of the microscale viscosity $\mu$ and heat
      conductivity $\kappa$ by their own Reynolds averages;
    \item Bernoulli's principle for the turbulent velocity
      fluctuations in \eqref{eq:bernoulli_p_u}.
  \end{enumerate}
\end{itemize}
In order to make the resulting system of equations suitable for
practical computations and modeling, two further parameterizations
need to be developed:
\begin{enumerate}
\item A closure of the hierarchy of the coarse-grained equations in
  \eqref{eq:coarse_grained_transport_equations}. Namely, we need to
  relate the coarse-grained skewness moment $\BS Q^*$ and the
  contracted fourth-order moment $\BS R^*$ to the coarse-grained
  density, pressure, stress and heat flux.
\item Suitable parameterizations of the coarse-grained collision
  operators $\BS C^*_S$ and $\BS c^*_q$, and the turbulent energy
  dissipation rate $\rho\varepsilon_t$.
\end{enumerate}
Below we elaborate on the first subject to the extent it allows to
avoid the second subject. The reason for this is that the collision
operators and the turbulent energy dissipation rate are the Reynolds
averages of nonlinear (with respect to the transported variables)
quantities, and thus should likely be modeled depending on a
particular application, with more detailed assumptions on the flow
properties put into place. Nonetheless, in Section
\ref{sec:computational_considerations} we suggest some crude
parameterizations for $\BS C^*_S$, $\BS c^*_q$ and
$\rho\varepsilon_t$.

\subsection{Grad closure for the coarse-grained transport equations}
\label{sec:Grad_closure}

Generally, there is little that can be concluded about the local
entropy state $S_l[\overline f]$ of $\overline f$. As Girimaji pointed
out in \cite{Gir}, the problem here is that $\overline f$ is the
statistical ensemble average of many realizations of $f$, where each
realization of $f$ generally has its own density $\rho$, velocity $\BS
u$, and pressure $p$ at a given spatial point $\BS x$. Therefore, even
if we assume that each $f$ in the statistical ensemble is the
corresponding Maxwellian state \eqref{eq:Maxwellian} with its own
$\rho$, $\BS u$ and $p$, the averaged sum of these states does not
have to be near a Maxwellian state. However, if $\overline f$ is far
from a Maxwellian, then both the Euler and the Navier-Stokes
approximations for the moments of such distribution become rather
questionable.

In contrast to the local entropy state, it turns out that the global
entropy state of the Reynolds-averaged Boltzmann equation behaves
similarly to the Reynolds average of the global entropy. In order to
better argue our point here, we first note that the Reynolds operator
satisfies Jensen's inequality.
\begin{proposition}[Jensen's inequality]
\label{prop:reynolds_jensen}
The Reynolds operator satisfies Jensen's inequality: for a convex
function $\phi$ and a function of the statistical ensemble $h$,
\begin{equation}
\overline{\phi(h)}\geq\phi(\overline h).
\end{equation}
\begin{proof}
Follows from the fact that the Reynolds operator is a statistical
average.
\end{proof}
\end{proposition}
This results in the following estimate for the entropy of the
Reynolds-averaged solution of the Boltzmann equation
\eqref{eq:boltzmann_equation}:
\begin{proposition}[Entropy of the Reynolds-averaged Boltzmann equation]
\label{prop:global_entropy}
Let $\overline f$ be the Reynolds-averaged distribution $f$ of the
Boltzmann equation in \eqref{eq:boltzmann_equation}, which apparently
satisfies
\begin{equation}
\parderiv{\overline f}t+\BS v\cdot\nabla_{\BS x}\overline f=
\overline{\coll(f)}.
\end{equation}
Then, both the local and global entropies of $\overline f$ are bounded
from below by the Reynolds averages of the local and global entropies
of $f$, respectively:
\begin{equation}
S_l[\overline f]\geq\overline{S_l[f]},\qquad
S_g[\overline f]\geq\overline{S_g[f]}.
\end{equation}
\begin{proof}
Observe that $\phi(f)=f\ln f$ is convex. Therefore, Jensen's
inequality yields
\begin{equation}
S_l[\overline f](t,\BS x)=-\int\overline f\ln\overline f\dif\BS v\geq
-\int\overline{f\ln f}\dif\BS v=\overline{S_l[f](t,\BS x)}.
\end{equation}
The corresponding inequality for the global entropy is obtained by
integrating in $\BS x$.
\end{proof}
\end{proposition}
The above result is not of much use for estimating the local behavior
of $\overline f$ at a spatial point $\BS x$; indeed, while the lower
bound estimate is valid, the upper bound for $S_l[\overline f]$ can be
different from the upper bound for $\overline{S_l[f]}$, due to the
fact that the local density, momentum and pressure are not preserved
between the members of the statistical ensemble (if they were, there
would likely be little need for the Reynolds averaging to begin
with). However, it is quite reasonable to assume that the members of
the statistical ensemble share the total mass, momentum and energy
constraints, because they have to be different realizations of the
same large-scale flow scenario, rather than being completely
unrelated. Therefore, the statistical ensemble members also share the
maximum global entropy state $\rho_M^{\text{g}}$ under these
constraints, which means that the maximum global entropy state is
invariant under the Reynolds averaging. As a result, one can show the
following inequality for the global relative entropy,
\begin{equation}
H_g[\overline f]=-S_g[\overline f]-\int\overline
f\ln\rho_M^{\text{g}}\dif\BS v\dif\BS x\leq-\overline{S_g[f]}
-\int \overline{f\ln\rho_M^{\text{g}}}\dif\BS v\dif\BS
x=\overline{H_g[f]}.
\end{equation}
The latter means that the global relative entropy of the
Reynolds-averaged distribution $\overline f$ decays at the same rate
as the Reynolds average of the ensemble global entropy states (that
is, $O(t^{-\infty})$, as shown by Desvillettes and Villani
\cite{DesVil}).

This situation is similar to what usually happens in the dynamics of
rarefied gases \cite{Gra,Gra2,Cer3,Lev,LevMor} where there is little time
scale separation between the growth rates of the local and global
entropies, and thus the higher-order moments do not rapidly converge
to their local equilibrium values, while still maintaining global
convergence on a slower time scale. Thus, the Grad closure
\cite{Gra,Gra2} appears to be a suitable option for closing the
transport equations in \eqref{eq:coarse_grained_transport_equations}
for the coarse-grained variables in
\eqref{eq:coarse_grained_variables}. Applying the Grad closure to the
coarse-grained moment transport equations in
\eqref{eq:coarse_grained_transport_equations} is identical to that of
the usual moment transport equations in
\eqref{eq:transport_equations}, as the Grad closure imposes no
restrictions on the structure of the collision terms, whether
Reynolds-averaged or not. The resulting system of equations for the
coarse-grained variables is the same as the one in
\eqref{eq:Grad_equations}, except that the collision terms are
different:
\begin{subequations}
\label{eq:coarse_grained_Grad_equations}
\begin{equation}
\label{eq:cg_Grad_density}
\parderiv{\rho^*}t+\div(\rho^*\BS u^*)=0,
\end{equation}
\begin{equation}
\parderiv{(\rho^*\BS u^*)}t+\div(\rho^*\BS u^*\otimes\BS u^*+p^*\BS I
+\BS S^*)=\BS 0,
\end{equation}
\begin{equation}
\label{eq:cg_Grad_pressure}
\parderiv{p^*}t+\div(p^*\BS u^*)+\frac 23\left[p^*\,\div\BS u^*
  +\BS S^*:(\nabla \otimes\BS u^*)+\div\BS q^*\right]=0,
\end{equation}
\begin{multline}
\label{eq:cg_Grad_stress}
\parderiv{\BS S^*}t+(\BS u^*\cdot\nabla)\BS S^* +\div(\BS u^*)
\BS S^*+2\symtr{\BS S^*(\nabla \otimes\BS u^*)} +\frac
45\symtr{\nabla\otimes\BS q^*}+\\+2p^*\symtr{\nabla\otimes\BS u^*
  }=\BS C^*_S,
\end{multline}
\begin{multline}
\label{eq:cg_Grad_heat_flux}
\parderiv{\BS q^*}t+\frac 75\big(\div(\BS q^*\otimes\BS u^*)+(\BS q^*
\cdot \nabla)\BS u^*\big)+\frac 25\big((\nabla\otimes\BS u^*)\BS
q^*-(\BS u^*\cdot\nabla)\BS q^*\big)-\\-\frac 1{\rho^*}\left(\frac
52 p^*\BS I+\BS S^* \right)\div(p^*\BS I+\BS S^*)+\div\left(
\frac {5p^{*2}}{2\rho^*}\BS I+\frac{7p^*}{2\rho^*} \BS S^*\right)
=\BS c^*_q.
\end{multline}
\end{subequations} 
The following arguments can be made in favor of the Grad closure in
\eqref{eq:coarse_grained_Grad_equations}:
\begin{itemize}
\item The ability to set the prescribed stress and heat flux. It
  cannot be assumed that the coarse-grained stress $\BS S^*$ and
  heat flux $\BS q^*$ are zero (or nearly zero), as there is not
  enough information about rapid Reynolds-averaged local entropy
  growth (or, equivalently, decay of $H_l[f]$) to substantiate
  that. Thus, a suitable moment closure must account for that. At the
  same time, the moments of such closure must be explicitly computable
  in terms of elementary functions, for the transport equations to
  have an explicit form. Also, one must remember that, due to
  Proposition \ref{prop:global_entropy}, the Reynolds average
  $\overline f$ has a global tendency to dissipate towards a normal
  distribution on a long time scale, and thus the closure should not
  be ``too far away'' from this state. The Grad closure meets all of
  these conditions: it is built around \eqref{eq:Maxwellian}, its
  moments are explicitly computable, and it also possesses the
  prescribed stress and heat flux, in addition to prescribed density,
  velocity and pressure.
\item Few restrictions for the parameterization of the coarse-grained
  collision terms. As the coarse-grained collision terms
  $\BS C^*_S$ and $\BS c^*_q$ are nonlinear and require separate
  (quite possibly empirical or semi-empirical) treatment, any need in
  prior assumptions on the structure of the collision terms (as in the
  Navier-Stokes approximation, for example) is undesirable.  In the
  Grad closure, there are no {\em a priori} assumptions on the form of
  collision terms, as they enter the equations as unspecified
  parameters.
\item Backward consistency with a weak/vanishing turbulence scenario.
  If the turbulence is weak or vanishing, then the coarse-grained
  variables in \eqref{eq:coarse_grained_variables} become the
  corresponding fine-grained variables in \eqref{eq:moments}. Then,
  the Grad equations for the coarse-grained variables naturally
  transition into the ordinary Grad equations in
  \eqref{eq:Grad_equations}, with appropriate collision terms.
\item Generalization of Millionschikov's hypothesis. In 1941,
  Millionschikov \cite{Mil,Mil2} asserted that the relations between
  the second and fourth turbulent statistical moments are related in
  approximately the same way as the corresponding moments of the
  normal distribution. Later in 1948, Heisenberg \cite{Hei}
  independently proposed the same hypothesis. Since then, some
  evidence was accumulated in favor of this hypothesis (see
  \cite{MonYag,MonYag2} and references therein). The Grad distribution
  generalizes Millionschikov's hypothesis, providing the explicit
  dependence of higher-order turbulent moments on the stress and heat
  flux in addition to the density, velocity and pressure.
\end{itemize}
However, there are also the following drawbacks:
\begin{itemize}
\item Increased computational complexity. Observe that the usual
  Navier-Stokes equations incorporate only five prognostic variables:
  the density, three velocity components, and pressure. For the Grad
  closure, the number of prognostic variables increases to 13 (plus
  one more for the turbulent energy $E_t$).
\item Somewhat questionable numerical stability. There are at least
  two types of possible numerical instabilities which can manifest in
  the Grad equations. The first one arises when the contribution of
  the collision terms for the stress and heat flux is very large, so
  that these variables are strongly damped, and thus the numerical
  stiffness (that is, oscillation of the numerical time-discretization
  polynomial) arises for high-accuracy integration schemes unless the
  time step is very small, or a specially tailored scheme, such as an
  implicit BDF, is used. The second numerical instability will arise
  if the collision terms $\BS C^*_S$ and $\BS c^*_q$ do not
  include diffusion, in which case the numerical solution will develop
  shock waves \cite{Gra} which will transfer the energy to high
  Fourier wavenumbers and cause the Gibbs oscillations. Below in
  Section \ref{sec:computational_considerations} we suggest an option
  to mitigate this problem.
\end{itemize}

\subsection{Coarse-grained viscosity and heat conductivity
approximations}

Observe that as the turbulence vanishes, the coarse-grained variables
in \eqref{eq:coarse_grained_variables} naturally become the
fine-grained variables in \eqref{eq:moments}, while the coarse-grained
transport equations in \eqref{eq:coarse_grained_transport_equations}
become the fine-grained transport equations in
\eqref{eq:transport_equations}. The latter suggests that, under the
conditions of sufficiently weak turbulence, the coarse-grained
collision terms $\BS C^*_S$ and $\BS c^*_q$ can be represented as
linear damping via the coarse-grained viscosity $\mu^*$ and heat
conductivity $\kappa^*$ as follows:
\begin{subequations}
\label{eq:cg_mu_kappa}
\begin{equation}
\BS C^*_S\approx-\frac{p^*}{\mu^*}\BS S^*,
\end{equation}
\begin{equation}
\BS c^*_q\approx-\frac 52\frac{p^*}{\kappa^*}\BS q^*,
\end{equation}
\end{subequations}
where $\mu^*$ and $\kappa^*$, of course, remain unspecified quantities
and must be somehow parameterized, possibly semi-empirically. However,
this situation is much simpler than what it was before from the
modeling perspective, as now two scalar quantities need to be
estimated, rather than two general collision operators. Below in
Section \ref{sec:computational_considerations} we suggest a crude
approximation for both $\mu^*$ and $\kappa^*$, which could serve as a
starting point in practical modeling, at least until the properties of
$\BS C^*_S$ and $\BS c^*_q$ become better studied.

The main drawback of the linear damping parameterization is the
following. In practice, the microscale viscosity and heat conductivity
are very small, while the turbulent fluctuations are large. In a
typical turbulent flow situation, $\BS S'\gg\BS S$ and $\BS
q'\gg\BS q$, such that the coarse-grained stress $\BS S^*$ and
heat flux $\BS q^*$ consist largely of the Reynolds-averaged turbulent
fluctuations,
\begin{equation}
\BS S^*\approx-\overline{\BS S'},\qquad\BS
q^*\approx-\overline{\BS q'}.
\end{equation}
In this setting, there is simply not enough information to determine
whether $\overline{p\BS S}$ and $\overline{p\BS q}$ (which are
very small quantities) are collinear, respectively, to
$p^*\BS S^*$ and $p^*\BS q^*$ (which are large quantities), as the
latter are determined largely by the statistical physics of the
turbulent flow, rather than microscale collisions. However, this
collinearity is a necessary condition to justify scalar quantities
$\mu^*$ and $\kappa^*$.

\subsection{Coarse-grained Navier-Stokes equations}

If, in addition to \eqref{eq:cg_mu_kappa}, the coarse-grained
viscosity $\mu^*$ and heat conductivity $\kappa^*$ are sufficiently
small, it enables the Navier-Stokes approximation of the form
\eqref{eq:simplified_stress_heat_flux}--\eqref{eq:Navier_Stokes_stress_heat_flux},
that is,
\begin{subequations}
\label{eq:coarse_grained_Navier_Stokes_stress_heat_flux}
\begin{equation}
\BS S^*=-2\mu^*\symtr{\nabla\otimes\BS u^*},
\end{equation}
\begin{equation}
\BS q^*=-\kappa^*\nabla(p^*/\rho),
\end{equation}
\end{subequations}
so that the coarse-grained Grad equations in
\eqref{eq:coarse_grained_Grad_equations} simplify to the
coarse-grained Navier-Stokes equations:
\begin{subequations}
\label{eq:coarse_grained_Navier_Stokes_equations}
\begin{equation}
\parderiv{\rho^*}t+\div(\rho^*\BS u^*)=0,
\end{equation}
\begin{equation}
\parderiv{(\rho^*\BS u^*)}t+\div\left(\rho^*\BS u^*\otimes\BS
u^*\right)+\nabla p^*=2\,\div\left(\mu^*\symtr{\nabla\otimes\BS
  u^*}\right),
\end{equation}
\begin{equation}
\parderiv{p^*}t+\div(p^*\BS u^*)+\frac 23 p^*\,\div\BS u^*=\frac 43
\mu^*\left\|\symtr{\nabla\otimes\BS u^*}\right\|^2+\frac
23\div( \kappa^*\nabla(p^*/\rho^*)),
\end{equation}
\end{subequations}
where, as in \eqref{eq:Navier_Stokes_equations}, one could consider
dropping the term with $\mu^*$ in the coarse-grained pressure
equation, since it acts as unbalanced forcing which tends to increase
$p^*$. In this situation, the coarse-grained forcing $F^*_t$ in the
turbulent energy transport equation
\eqref{eq:turbulent_energy_equation} becomes
\begin{multline}
\label{eq:F_t_Navier_Stokes}
F^*_t=\left(\frac 12\nabla-\frac{\nabla\rho}\rho\right)\cdot \left(
(\kappa-\kappa^*)\nabla\left(\frac{p^*}\rho\right)\right)-\\-2(\mu-
\mu^*)\left( \nabla\otimes\BS u^*+\frac\mu\rho(\nabla\otimes\nabla)
+\frac\mu{ \rho^2} \big((\nabla\otimes\nabla)\rho\big)-\frac{5\mu}{3
  \rho^3}( \nabla\rho \otimes\nabla \rho)\right):\symtr{\nabla\otimes
  \BS u^*}+\\+\left(3\nabla-2\frac{\nabla\rho} \rho\right) \cdot
\left(\left(\frac\mu\rho\div+\frac{2\mu}{3\rho^2} \nabla\rho \right)
\left((\mu-\mu^*)\symtr{\nabla\otimes \BS u^* }\right) \right).
\end{multline}
Assuming that the coarse-grained viscosity $\mu^*$ and heat
conductivity $\kappa^*$ are large enough in comparison to the
microscale viscosity $\mu$ and heat conductivity $\kappa$ (but at the
same time still small enough to enable the Navier-Stokes
approximations in
\eqref{eq:coarse_grained_Navier_Stokes_stress_heat_flux}), one can
simplify the above expression as
\begin{equation}
F^*_t=-\left(\frac 12\nabla-\frac{\nabla\rho}\rho\right)\cdot\left(
\kappa^*\nabla\left(\frac{p^*}\rho\right)\right)+2\mu^*\left\|\symtr{
  \nabla \otimes\BS u^*} \right\|^2.
\end{equation}
The main advantage of the coarse-grained Navier-Stokes equations in
\eqref{eq:coarse_grained_Navier_Stokes_equations} is that they are
technically not very different from the usual Navier-Stokes equations,
for which many computational methods have already been
developed. Here, we have the additional transport equation for the
turbulent energy transport in \eqref{eq:turbulent_energy_equation},
but, again, it is not much different from the usual pressure transport
equation, except that it has the additional forcing, dissipation, and
diffusion.

The main drawback of the coarse-grained Navier-Stokes approximation
is, however, the following. Even if linear damping
\eqref{eq:cg_mu_kappa} applies for a given type of turbulence (which
does not restrict the use of the coarse-grained Grad equations in
\eqref{eq:coarse_grained_Grad_equations}), the necessary condition for
the coarse-grained Navier-Stokes approximation in
\eqref{eq:coarse_grained_Navier_Stokes_equations} to be viable is that
the evolution time scale of the coarse-grained stress and heat flux
must be much faster than that for the density, velocity and
pressure. This condition requires that both the coarse-grained
viscosity and heat conductivity must be very small (which is usually
the case for the conventional microscale $\mu$ and $\kappa$). However,
often in the models with the turbulent viscosity approximation these
parameters can be several orders of magnitude larger than the
microscale viscosity. In this situation, the time-scale separation is
unlikely to exist, and thus the turbulent Navier-Stokes approximation
may not apply.

\subsection{Coarse-grained Burnett and super-Burnett equations}

A key advantage of the new coarse-grained variables in
\eqref{eq:coarse_grained_variables} is that they obey the well-studied
hierarchy of the moment equations in
\eqref{eq:coarse_grained_transport_equations}. Because of this, one
has an opportunity to use existing closure methods for these
equations. Above we showed how to derive the coarse-grained Grad
\eqref{eq:coarse_grained_Grad_equations} and Navier-Stokes
\eqref{eq:coarse_grained_Navier_Stokes_equations} closures for this
hierarchy, but one does not have to stop there. Using the ratio
$\mu^*/p^*$ (or, equivalently, $\kappa^*/p^*$) as a small parameter
akin to the usual Knudsen number in the molecular kinetics, one can
carry out the standard Chapman-Enskog perturbation expansion into the
higher orders, obtaining the coarse-grained Burnett \cite{ChaCow} and
super-Burnett \cite{Shav} equations in the same way they are derived
for the conventional transport equations. While these equations are
generally ill-posed, some relaxation-type regularizations were
developed for them \cite{JinSle}.  Here we do not elaborate on this
further, as the derivation itself is straightforward, while a more
detailed study of the resulting coarse-grained Burnett or
super-Burnett equations is an entirely separate topic. With the
higher-order Chapman-Enskog expansion, the expressions for
$\BS S^*$ and $\BS q^*$ in
\eqref{eq:coarse_grained_Navier_Stokes_stress_heat_flux} will contain
additional terms, scaled by powers of $\mu^*/p^*$ and $\kappa^*/p^*$,
which will accordingly affect the expression for $F^*_t$ in
\eqref{eq:F_t_Navier_Stokes}. The transport and dissipation parts of
the equation for $E_t$ in \eqref{eq:turbulent_energy_equation} will
remain the same.

\section{Practical considerations for a computational implementation}
\label{sec:computational_considerations}

In this section we discuss possible issues arising in the course of
numerical implementation and modeling of the transport equations,
developed above.

\subsection{Inclusion of the coarse-grained spatial scale information}

A key subject which was thus far left out of the picture is the
spatial scale information about the coarse-grained and turbulent
scales. While this scale information can be avoided in the formal
coarse-grained transport formulation presented above (simply due to
the fact that the Boltzmann equation is linear except for its
collision operator), it naturally resurfaces when one has to
discretize the transport equations for a numerical simulation. For
example, if the discretization scale is so fine that it resolves the
viscous molecular dissipation, then, clearly, the turbulent energy
$E_t$ must be (nearly) zero, as should be the coarse-grained stress
$\BS S^*$ and heat flux $\BS q^*$.  On the other hand, if one
chooses to coarsen the spatial discretization mesh, then $E_t$,
$\BS S^*$ and $\BS q^*$ should naturally become larger, which
means that their corresponding dissipation terms must be weakened
appropriately. The only way to set these conditions in the transport
equations is to adjust the turbulent energy dissipation rate
$\rho\varepsilon_t$ and damping in the coarse-grained collision terms
$\BS C^*_S$ and $\BS c^*_q$ (or, alternatively, coarse-grained
viscosity $\mu^*$ and heat conductivity $\kappa^*$), solely because
there is nothing else in the transport equations that can be changed.

At first, this situation appears to conflict with the way the Reynolds
average is initially set up as a statistical average, since the
ensemble averaging is formally unrelated to the spatial filtering
(like the one in the LES). To clarify the situation, here we offer an
informal explanation how the spatial scaling information can be
incorporated into the Reynolds averaging.  The Reynolds averaging
process consists of two stages -- first, ``ensemble generation'', and,
second, ``ensemble averaging'' (as purely a thought process, of
course, since there is no explicit numerical ensemble generation or
averaging involved). Here we claim that the spatial scale information
is taken into account during the ``ensemble generation''
stage. Indeed, observe that all ensemble members must share the
coarse-grained scale features (as it is the ``resolved'' scale),
while, of course, differing on the turbulent scale. Therefore, if the
coarse-grained scale is in fact so fine that it resolves even the
molecular viscosity effects, the corresponding statistical ensemble
must, essentially, consist of a single member (or many identical
members), as any difference between the ensemble members on the
unresolved scale is nearly instantaneously damped to zero by the
molecular viscosity. Conversely, if the coarse-grained scale is not
too fine, then it lifts the difference between the ensemble members
into larger spatial scales, and hence weakens their dissipation.

This leads to the natural conclusion that the spatial scale
information must be encoded into the dissipative terms parameterizing
the turbulent energy decay rate $\rho\varepsilon_t$ and the
coarse-grained collision terms $\BS C^*_S$ and
$\BS c^*_q$. Particularly, in the limit as the coarse-grained scale
becomes fine enough to resolve the molecular viscosity scale, the
turbulent energy decay rate $\rho\varepsilon_t$ must be strong enough
to ensure that the turbulent energy $E_t$ is (almost) zero, while the
coarse-grained collision terms $\BS C^*_S$ and $\BS c^*_q$ must
transform into their fine-grained analogs in
\eqref{eq:Navier_Stokes_collisions}.

\subsection{Numerical stability and possible shock formation}

From was is developed above, one can envision three general scenarios
of a practical computational set-up:
\begin{enumerate}
\item The coarse-grained Grad transport equations
  \eqref{eq:coarse_grained_Grad_equations} with the specific,
  problem-dependent parameterization of the collision terms;
\item The coarse-grained Navier-Stokes transport equations
  \eqref{eq:coarse_grained_Navier_Stokes_equations} with the
  coarse-grained viscosity and heat conductivity;
\item The coarse-grained Grad transport equations
  \eqref{eq:coarse_grained_Grad_equations} with collision terms
  parameterized via linear damping as in \eqref{eq:cg_mu_kappa}, with
  coarse-grained viscosity and heat conductivity.
\end{enumerate}
There is not much to comment on in the first two scenarios, because
the latter is comprised by the well-studied Navier-Stokes equations,
while the computational properties of the former heavily depend on the
implementation of the coarse-grained collision terms. The third
scenario is, however, of more interest, since it is typically used in
rarefied gas dynamics \cite{Gra,Gra2,Stru,StruTor,TorStru}. In the
turbulent flow simulations, the third scenario is likely to appear
when the turbulence is still weak enough to parameterize the
collisions by linear damping \eqref{eq:cg_mu_kappa}, but at the same
time the coarse-grained viscosity $\mu^*$ and heat conductivity
$\kappa^*$ are not small enough to enable the Navier-Stokes
approximations of $\BS S^*$ and $\BS q^*$ via
\eqref{eq:coarse_grained_Navier_Stokes_stress_heat_flux}. In this
situation, observe that the resulting transport equations for the
coarse-grained variables do not have any diffusion terms, which leads
to the creation of shocks at high Mach numbers \cite{Gra}.

In this scenario, one might consider the regularization strategy
proposed by Struchtrup and Torrilhon \cite{StruTor,TorStru}, which is,
roughly put, a type of the Chapman-Enskog expansion built around the
Grad state in \eqref{eq:grad_state}, and applied to the transport
equations for $\BS Q^*$ and $\BS R^*$. The way this regularization
works is the same as in the Navier-Stokes parameterizations in
\eqref{eq:simplified_stress_heat_flux}--\eqref{eq:Navier_Stokes_stress_heat_flux},
yielding additional diffusion terms in the transport equations for the
coarse-grained stress \eqref{eq:cg_Grad_stress} and heat flux
\eqref{eq:cg_Grad_heat_flux}, and thus dissipating the shock
formation. The diffusion from these extra terms manifests on the scale
$\mu^*/p^*$ (or, alternatively, $\kappa^*/p^*$), and extends to the
larger scales as the coarse-grained viscosity and heat conductivity
increase, thus allowing for coarser computational discretization.

\subsection{A crude approximation for the turbulent energy dissipation
rate}

The simplest way to model the turbulent energy dissipation rate
$\rho\varepsilon_t$ is by using the dimensional analysis. Above, we
concluded that a parameterization for $\rho\varepsilon_t$ should
depend on a length scale parameter which characterizes the
coarse-grained spatial scale (such as the spatial discretization
size). Let us now assume that $\varepsilon_t$ may {\em only} depend on
such a spatial scale parameter $L$, and the turbulent energy $E_t$
itself, scaled by the density $\rho$:
\begin{equation}
\varepsilon_t\propto \left(E_t/\rho\right)^\alpha L^\beta,
\end{equation}
where the constants $\alpha$ and $\beta$ must be determined by
considering physical dimensions. Now, observe that the dimension of
$E_t/\rho$ is length$^2/$time$^2$, while the dimension of
$\varepsilon_t$ is length$^2/$time$^3$. Thus, the only way to equalize
dimensions in the expression above is to set $\alpha=3/2$,
$\beta=-1$, which yields
\begin{equation}
\label{eq:rho_e_taylor}
\varepsilon_t\propto\frac{E_t^{3/2}}{\rho^{3/2}L},\qquad
\rho\varepsilon_t\propto\frac{E_t^{3/2}}{\rho^{1/2}L}.
\end{equation}
This well-known simple parameterization was proposed by Taylor
\cite{Tay}. Observe that it is in agreement with what was concluded
earlier; namely, it disappears when $E_t\to 0$ (and thus does not let
$E_t$ become negative), and increases for fixed $E_t$ as $L\to 0$,
thus more strongly driving $E_t$ towards zero in the case the
coarse-grained spatial resolution scale becomes more refined.

However, we can immediately see that the simple parameterization above
has a serious drawback. Recall that we agreed in Section
\ref{sec:coarse_grained} that the new coarse-grained transport
equations must have solutions which do not exhibit rapid oscillations
at small scales, otherwise the very point of introducing the
coarse-grained equations would become moot. Nonetheless, with the
turbulent energy decay rate $\rho\varepsilon_t$ given as above in
\eqref{eq:rho_e_taylor}, the turbulent energy transport equation in
\eqref{eq:turbulent_energy_equation} has no diffusive terms of
sufficient strength (the existing diffusion term there manifests on
the molecular scale), and in that respect the equation in
\eqref{eq:turbulent_energy_equation} is no different from the
conventional transport equations in \eqref{eq:transport_equations} we
started with.

Therefore, here we contend that the turbulent energy decay rate
$\rho\varepsilon_t$ must incorporate a diffusion term which manifests
on the coarse-grained scale. In particular, we propose a simple
extension of \eqref{eq:rho_e_taylor} based, again, on a dimensional
argument:
\begin{equation}
\label{eq:rho_e}
\rho\varepsilon_t\propto\left(\frac{E_t}\rho\right)^{1/2}\left(\frac{E_t}L
+\alpha L\Delta E_t\right),
\end{equation}
where $\alpha$ is a scalar non-dimensional parameter. Observe that
above the first term is exactly the Taylor parameterization from
\eqref{eq:rho_e_taylor}. The additional term, scaled by $\alpha$, is a
diffusion term, which, contrary to the linear damping term, increases
when $L$ becomes large, extending the diffusion scale proportionally
to the coarse-grained scale parameter $L$. Conversely, the diffusion
term disappears as the coarse-grained scale becomes more refined. This
seems to be in consistence with what one would expect from the
behavior of the turbulent energy.

Of course, the parameterization in \eqref{eq:rho_e} is the simplest
model for $\rho\varepsilon_t$ one could possibly come up with, but at
least it seems to be a reasonable starting point for use in
computational modeling. There also exist more sophisticated models for
$\varepsilon_t$ (such as, for example, the $k$-$\varepsilon$ and
$k$-$\omega$ models \cite{Wilc}) but they require a separate transport
equation for $\varepsilon_t$, which, in turn, introduces additional
closure problems via unspecified parameters.

\subsection{Crude approximations for the coarse-grained collision terms}

Here we offer a very crude parameterization of the coarse-grained
collision terms in the form of a linear damping, which, of course,
will not stand any critique when the properties of the coarse-grained
collision terms become better studied, but at present may at least
serve as a starting point in the computational modeling. Here we
assume that the coarse-grained collision terms $\BS C^*_S$ and
$\BS c^*_q$ can be modeled by a linear damping of the form
\eqref{eq:cg_mu_kappa}, where the coarse-grained viscosity $\mu^*$ and
heat conductivity $\kappa^*$ are to be approximated somehow. From the
prior considerations, we expect that $\mu^*$ and $\kappa^*$ must be
proportional to the coarse-grained length scale (that is, they must
become small when the length scale is short, and vice versa). At the
same time, we know that, as the coarse-grained length scale becomes
short, $\mu^*$ and $\kappa^*$ must become their microscale
counterparts $\mu$ and $\kappa$, respectively. The form of the
microscale $\mu$ and $\kappa$ is given in Grad \cite{Gra} as
\begin{equation}
\mu,\kappa\propto\frac m{\sigma^2}\left(\frac p\rho\right)^{1/2}=\frac
m{\sigma^3}\sigma\left(\frac p\rho\right)^{1/2}=\rho_m\sigma \left(
\frac p\rho\right)^{1/2},
\end{equation}
where $m$ is the mass of the molecule, $\sigma$ is its linear size
(diameter or radius), and $\rho_m$ is the ``density'' of a
molecule. Now, with help of a vivid imagination, one could interpret
the collision processes on a coarse-grained scale as ones with rather
large and soft molecules. This means that for the coarse-grained
viscosity or heat conductivity one likely has to keep the ``density''
$\rho_m$ fixed, but increase the ``size'' of the molecule to be that
of the characteristic length of the coarse-grained scale, while at the
same time also parameterizing $p$ via either its coarse-grained
counterpart, or its Reynolds average:
\begin{equation}
\label{eq:cg_param_mu_kappa}
\mu^*,\kappa^*\propto\left(\frac p\rho\right)^{1/2}L\propto\left\{
\begin{array}{l}\displaystyle\left(\frac{p^*}\rho\right)^{1/2}L,\quad
\text{or}\\ \displaystyle\left(\frac{\overline p}\rho\right)^{1/2}L
=\left(\frac{3p^*-2E_t}{3\rho}\right)^{1/2}L,\end{array}\right.
\end{equation}
where the choice is between the lack of feedback from the small
scales, or a feedback which damps the viscosity and heat conductivity
if the turbulent energy becomes too large. The coefficients of
proportionality should likely be chosen so that in the limit as the
resolved scale becomes the molecular viscosity scale, both $\mu^*$ and
$\kappa^*$ become their microscale counterparts. In fact, they do not
necessarily have to be constants, as that would impose the same
Prandtl number on the coarse-grained scale, which may not necessarily
be the case in practice.

The relation above constitutes the simplest, crudest parameterization
for the coarse-grained viscosity and heat conductivity. This
parameterization is, however, consistent with what is expected of it,
namely, it increases the coarse-grained viscosity and heat
conductivity in proportion to the coarse-grained spatial scale, and,
if the turbulent energy feedback option is chosen, it also tends to
decrease them as $E_t$ increases, thus providing an additional
balancing mechanism. Note, however, that a regularization along the
lines of Struchtrup and Torrilhon \cite{StruTor,TorStru} is needed for
a linear damping parameterization in \eqref{eq:cg_param_mu_kappa} to
improve the numerical stability of the coarse-grained Grad equations
in \eqref{eq:coarse_grained_Grad_equations}. Alternatively, if the
resulting $\mu^*$ and $\kappa^*$ are found to be small enough to allow
the steady-state parameterization of the kind in
\eqref{eq:coarse_grained_Navier_Stokes_stress_heat_flux}, then one can
optionally use the coarse-grained Navier-Stokes equations
\eqref{eq:coarse_grained_Navier_Stokes_equations} instead of the
coarse-grained Grad equations \eqref{eq:coarse_grained_Grad_equations}
(or suitably regularized Burnett/super-Burnett expansions, for better
accuracy with not-so-small $\mu^*$ and $\kappa^*$).

\section{Summary and future research}
\label{sec:summary}

In this work, we propose a new framework for the coarse-grained
transport of a turbulent flow, which is based on the Reynolds
averaging of the Boltzmann equation in \eqref{eq:boltzmann_equation},
rather than the conventional Reynolds averaging of the Navier-Stokes
equations in \eqref{eq:Navier_Stokes_equations}. The proposed
framework consists of the equations in
\eqref{eq:coarse_grained_transport_equations} for the transport of the
coarse-grained variables of the flow in
\eqref{eq:coarse_grained_variables}, and the equation for the
transport of the turbulent energy in
\eqref{eq:turbulent_energy_equation} under the approximation of the
anelastic turbulence. We also propose two different closures for the
hierarchy of the transport equations for the coarse-grained variables:
the more general Grad closure in
\eqref{eq:coarse_grained_Grad_equations}, and the more simple
Navier-Stokes closure formulation in
\eqref{eq:coarse_grained_Navier_Stokes_equations}, which can likely be
expanded into the higher Burnett or super-Burnett orders to improve
accuracy. The proposed transport model includes three unknown
parameters: the turbulent energy dissipation rate $\rho\varepsilon_t$
in the turbulent energy equation \eqref{eq:turbulent_energy_equation},
the coarse-grained stress collision operator $\BS C^*_S$ (or,
equivalently, the corresponding coarse-grained viscosity $\mu^*$) in
the coarse-grained stress equation \eqref{eq:cg_Grad_stress}, and the
coarse-grained heat flux collision operator $\BS c^*_q$ (or,
equivalently, the corresponding coarse-grained heat conductivity
$\kappa^*$) in the coarse-grained heat flux equation
\eqref{eq:cg_Grad_heat_flux}. These three parameters cannot be defined
exactly within the scope of the new transport model, as they depend on
the properties of the nonlinear fluid-fluid interactions, and require
separate treatment. We also suggest crude parameterizations for
$\rho\varepsilon_t$, $\BS C^*_S$ and $\BS c^*_q$ based on
dimensional analysis, to be used as a starting point in the practical
computational modeling.

{\bf Future research.} The main deficiency of the developed
coarse-grained transport framework is that it is derived under the
assumption of a monatomic ideal gas model. While this assumption
greatly simplifies calculations, at the same time it severely
restricts the applicability of the framework, as most gases in the
surrounding nature are at least diatomic (e.g. the air). At the same
time, such a framework would likely be of use in the applications
which involve the circulation of the large scale atmosphere, as it is
well known that the atmosphere is a highly turbulent medium, and at
the same time present computational limitations restrict the spatial
resolution of the global circulation models to rather coarse
meshes. Thus, the development of the analogous coarse-grained
transport framework for the polyatomic (or at least diatomic) ideal
gases is presently our main priority.

\subsection*{Acknowledgment.}

The author thanks Ibrahim Fatkullin and Roman Shvydkoy for interesting
discussions. The author also thanks Francis Filbet, Cl\'ement Mouhot,
Lorenzo Pareschi and the Society for Industrial and Applied
Mathematics for their permission to adapt a figure from
\cite{FilMouPar}. The work was supported by the National Science
Foundation CAREER grant DMS-0845760, and the Office of Naval Research
grant N00014-15-1-2036.

\end{document}